\documentclass[prd,onecolumn,notitlepage,showpacs,preprintnumbers,amsmath,amssymb,nofootinbib,APS,10pt,superscriptaddress]{revtex4-1}

\usepackage{dcolumn}
\usepackage{bm}
\usepackage[applemac]{inputenc}
\usepackage[spanish,english]{babel}
\usepackage{amsmath,amssymb,amsfonts,latexsym,cancel}
\usepackage{graphicx} 
\usepackage{color}
\usepackage{soul}
\usepackage{ulem}
\usepackage{hyperref}
\usepackage{slashed}

\allowdisplaybreaks

\newcommand{\be}{\begin{equation}}
\newcommand{\ee}{\end{equation}}
\newcommand{\bea}{\begin{eqnarray}}
\newcommand{\eea}{\end{eqnarray}}

\begin{document}

\title{{\bf Chiral fermion anomaly as a memory effect}}

%

%
\author{Adri\'an del R\'io}\email{adrian.rio@uv.es}
\affiliation{Departamento de F\'isica Te\'orica and IFIC, Universitat de Valencia - CSIC.  \\ Dr. Moliner 50, 46100, Burjassot (Valencia), Spain.}
\affiliation{Institute for Gravitation and the Cosmos \& Physics Department,The Pennsylvania State University, University Park, PA 16802 USA}
\author{Ivan Agullo}\email{agullo@lsu.edu}
\affiliation{Department of Physics and Astronomy, Louisiana State University, Baton Rouge, LA 70803-4001, USA}
\affiliation{Perimeter Institute for Theoretical Physics, 31 Caroline Street North, Waterloo, Ontario, Canada N2L 2Y5}

\begin{abstract}
We study the non-conservation of the chiral charge of  Dirac fields  between past and future null infinity due to the Adler-Bell-Jackiw chiral anomaly. In previous investigations \cite{dR21}, we found that this charge fails to be conserved if electromagnetic sources in the bulk emit circularly polarized radiation. In this article, we unravel yet another contribution  coming from the non-zero, infrared  ``soft'' charges of the external, electromagnetic field.
This new contribution can be interpreted as another manifestation of the ordinary memory effect produced by transitions between different infrared  sectors of Maxwell theory,  but now on test quantum fields rather than on test classical particles. In other words,  a flux of electromagnetic waves can leave a memory on quantum fermion states in the form of a permanent, net helicity. We elaborate this idea  in both $1+1$ and $3+1$ dimensions. We also show that, in sharp contrast, gravitational infrared charges do not contribute to the fermion chiral anomaly.
 \end{abstract}

\date{\today}
\maketitle


\section{Introduction}
\label{introduction}

Not all symmetries of a classical theory remain exact after quantization. When this occurs, i.e., when a symmetry of the action is broken by quantum effects, one speaks about anomalies \cite{B96}. Anomalies were first discovered in the late 1960s, in the seminal works by Adler, Bell and Jackiw, as an attempt of solving the pion decay puzzle \cite{A69, BJ69}. They found that the chiral symmetry of the action of a massless Dirac field $\Psi(x)$ that interacts with  an electromagnetic   background is broken in quantum field theory. Mathematically, this outstanding result is beautifully encoded in the non-conservation equation of the fermionic chiral current $j^{a}_A(x)=\bar \Psi(x) \gamma^{a} \gamma^5\Psi(x)$, which on a 3+1 dimensional Minkowski spacetime takes the form
 \be  \label{emanomaly} \left< \nabla_{a}j^{a}_A\right>= -\frac{\hbar q^2}{8\pi^2} F_{ab}\, ^{\star}F^{ab}\, , \ee
where  $F_{ab}$ is the field strength of the background electromagnetic field, $^{\star}F^{ab}$ its Hodge dual,  and $q$ the charge of the fermion. This is the celebrated chiral or axial anomaly. 

Besides electromagnetic fields, gravitational backgrounds have also the ability of triggering a chiral anomaly,  as it was soon after found in   \cite{K69,DS72,EF76}. Mathematically, this contribution generalizes the previous equation by adding a new term proportional to the pseudo-scalar curvature invariant $R_{abcd}\, ^{\star}R^{abcd}$, where $R_{abcd}$  is the Riemann tensor and ${^*R}_{abcd}$ its Hodge dual with respect to the first two indices:
\be \label{totalanomaly} \left< \nabla_{a}j^{a}_A\right>= -\frac{\hbar q^2}{8\pi^2} F_{ab}\, ^{\star}F^{ab}+\frac{\hbar}{192\pi^2} R_{abcd}\, ^{\star}R^{abcd}\, .\ee
The following years experienced an  outbreak of fascinating results involving anomalies, both regarding physics and mathematics. Examples include,  besides the prediction of the neutral pion decay rate to two photons, applications to the matter-antimatter asymmetry of the universe, the U(1) and strong CP problems in QCD, implications for anomaly cancellation in the Standard Model  (see \cite{Schwartz2014} for a nice summary of all these applications), and connections with the index theorems in geometric analysis \cite{EguchiHansonGilkey1980, BastianelliVanNieuwenhuizen2006, Nakahara}. The notion of chiral anomalies has also been extended to other fields, including integer spin fields \cite{AdRNS16, AdRNS17, AdRNS18a, AdRNS18b}. 

In this article, we investigate yet another aspect of chiral anomalies, related to {\it global} properties of both the fermionic and the background fields. These global properties appear when discussing the Noether charge associated with the chiral current, namely $Q_A=\int_{\Sigma}\, d\Sigma\,  j^{0}_A$, where the integral is computed on any constant-time Cauchy hypersurface $\Sigma$. For the sake of clarity, let us focus on electromagnetic backgrounds, although  similar arguments apply also for gravitational backgrounds ---except for some important differences that we unravel  in this article. Classically, the chiral charge $Q_A$ of a Dirac field measures the difference in the amplitude of the two helical components of the field. Quantum mechanically, this quantity translates to the difference in the number of positive and negative helicity particles, together with possible contributions from ``vacuum polarization''. The charge $Q_A$ is strictly conserved in the classical theory, but it is not quantum mechanically due to non-conservation of the current \eqref{totalanomaly}. This can be easily shown by considering any two Cauchy hypersurfaces, $\Sigma_{\rm in}$ and $\Sigma_{\rm out}$, and noticing that the change in the vacuum expectation value of $Q_A$  between  $\Sigma_{\rm in}$ and $\Sigma_{\rm out}$ is equal to the integral of  $\left< \nabla_{a}j^{a}_A\right>$ in the four dimensional region $R$ bounded by $\Sigma_{\rm in}$ and $\Sigma_{\rm out}:$
\be \int_R  d^4x\, \left< \nabla_{\mu}j^{\mu}_A\right>=\int_R d^4x \,  \big(\langle  \partial_t j^{0}_A\rangle+ \langle \vec \nabla \cdot \vec j_A\rangle\big)=\int_R d^4x \,  \langle  \partial_t j^{0}_A\rangle=\left< Q_A\right>_{\Sigma_{\rm out}}- \left< Q_A \right>_{\Sigma_{\rm in}}\, , \label{integral}
\ee
where, in the second equality, the second term vanishes due to Stoke's theorem and standard fall-off conditions of the fields at spatial infinity. Hence, a non-zero value of the integral $\int_R  d^4x \, \left< \nabla_{a}j^{a}_A\right>$ implies the non conservation of the chiral charge of  the quantum fermionic field.

Our goal is to understand what characteristics the electromagnetic  backgrounds must have to  produce a non-zero value of this integral. It is easy to check  that the pseudo-scalar $F_{ab}\, ^{\star}F^{ab}$ appearing in  \eqref{emanomaly} can be written as the divergence of a vector, $j_{\rm CS}^{a}=2 A_{b}\, ^{\star}F^{ab}$, where  $A_{a}$ is  the electromagnetic potential ---CS stands for Chern-Simons. Repeating the steps used to produce Eqn.~\eqref{integral},  Eqns.~\eqref{emanomaly} and (\ref{integral})   automatically imply that the fermionic chiral charge $\left< Q_A\right>$  fails to be conserved if and only if the scalar $Q_{\rm CS}=\int_{\Sigma} d\Sigma\, j_{CS}^0$ associated  with the vector $j_{\rm CS}^{a}$ changes between $\Sigma_{\rm in}$ and $\Sigma_{\rm out}$:
\be \label{aaa} \left< Q_A\right>_{\Sigma_{\rm out}}- \left< Q_A \right>_{\Sigma_{\rm in}}=-\frac{q^2\hbar }{8 \pi^2} \,\left[ Q_{CS}|_{\Sigma_{\rm out}} -  Q_{CS}|_{\Sigma_{\rm in}}\right] \, , \ee
This simple observation provides an interesting strategy to classify  electromagnetic backgrounds that are able to trigger an anomalous non-conservation of the chiral charge of fermionic quantum fields propagating thereon. 
\footnote{For Yang-Mills  fields, the standard strategy consists in looking for instanton solutions in Euclidean space. However, there are no instanton solutions of Maxwell (Abelian) equations in 4 dimensions. Furthermore, it is useful to work directly within the framework of asymptotically flat spacetimes, since it captures the full causal structure of physical (Lorentzian) spacetimes.}
This strategy was initiated in \cite{dRetal2020, dR21} for both electromagnetic and gravitational backgrounds, where some aspects of the scalar $Q_{\rm CS}$ were analyzed for asymptotically flat spacetimes, in which the hypersurfaces $\Sigma_{\rm in}$ and $\Sigma_{\rm out}$ can be chosen to be past ($\mathcal{I}^-$) and future null infinity ($\mathcal{I}^+$), respectively. This is a natural choice when studying massless quantum fields.
It was  shown that, at these limiting surfaces, the scalar $Q_{\rm CS}$ receives a contribution from the net helicity of the radiative content of the electromagnetic and gravitational fields. This implies that, if there are  sources in the bulk emitting helical or circularly polarized radiation ---in gravity, this  happens, for instance, in the coalescence of a large family of binary black hole mergers \cite{dRetal2020, SGdR23}--- there is a net change of $Q_{\rm CS}$ between $\mathcal{I}^-$ and $\mathcal{I}^+$, which  induces a change in the chiral charge  $\langle Q_A\rangle$ of fermionic quantum fields propagating thereon. This is a profound relation between the radiative content of the background field and the chiral charge of  quantum fields. We emphasize that this is a quantum effect; classically, $\langle Q_A\rangle$ is strictly conserved for massless fields, regardless of the properties of the electromagnetic and gravitational backgrounds.

This article unravels another contribution to $Q_{\rm CS}$ ---which, consequently, also  acts  as a source of  fermionic helicity $\langle Q_A\rangle$--- originated in the existence of certain electromagnetic infrared, or ``soft'' charges. Infrared charges have received a good deal of attention in the recent past, due to their theoretical importance in the study of the $S$-matrix in quantum electrodynamics  and quantum gravity, and due to their connection with soft theorems (see the reviews \cite{ACL18, S17} and references therein).  On the other hand, non-zero infrared charges indicate the generation of ``memory effects'' in physical systems.  To give an example, test charged particles can experience a permanent change in their  velocity (a ``kick'') after the passage of electromagnetic waves \cite{BG13}. In electrodynamics, this was the first  example of memory  reported in the literature. Other memory effects have been  identified in recent  years (see for instance \cite{OS23, SO21} for an effect related  with the helicity of radiation, and references therein).

Therefore, the results of this article can be interpreted as another type of memory effect produced by  infrared charges, now on {\it test quantum fields} rather than on classical test particles. Quite interestingly, we find that this new manifestation of electromagnetic memory effect does not occur for the gravitationally-induced chiral anomaly. 

The rest of this article is organized as follows. Section \ref{sec2} introduces a simple example of  pedagogical value:  a massless Dirac field in 1+1-dimensional flat spacetime coupled to an electromagnetic background. Section \ref{sec3} contains a brief summary of the asymptotic properties of the electromagnetic field at past and future null infinity, including the notion of soft charges and memory effect in this framework; readers already familiar with the notation can skip this section. Section \ref{sec4} contains the main analysis of this article, where the contribution of  soft electromagnetic charges  for the Adler-Bell-Jackiw anomaly is derived.  This section also includes a simple  example of an electromagnetic configuration for which the relevant infrared charges are different from zero. The gravitational case is discussed in section \ref{sec5}, and section \ref{sec6} closes the paper with a few conclusions and remarks.

Throughout this paper, we use  geometric units in which $G=c=1$, and we keep $\hbar$ explicit in our equations to emphasize quantum effects. The metric signature is chosen to be $(-,+,+,+)$; $\nabla_a$  represents the  Levi-Civita connection; the Riemann tensor is defined by\, $2 \nabla_{[a}\,\nabla_{b]}v_c=:R_{abc}{}^{d}v_d$\, for any covector $v_d$. Unless otherwise stated, all tensors will be assumed to be smooth.

\section{Chiral anomaly in two dimensions} \label{sec2}

This section discusses a massless, charged  Dirac field in a 1+1-dimensional flat spacetime,  propagating on a homogeneous electric background with finite support in time. To make our arguments simpler, we assume  that the spacetime manifold is  $\mathbb{R}\times \mathbb{S}^1$, i.e.,  the spatial dimension has been compactified to the circle.  The electric field is assumed to be strong enough to make the backreaction of the quantum field negligibly small.  This  setup has great pedagogical value to illustrate some of the main messages of this paper, particularly  the relation between the chiral fermion anomaly and the memory effect. We also discuss the relation of these concepts with spontaneous quantum particle-pair creation. 

In a 1+1-dimensional spacetime, $F_{ab}$ has only one independent component ---the electric field--- $F_{ab}=E\, \epsilon_{ab}$, where $\epsilon_{ab}$ is the totally anti-symmetric tensor.  Among the two Maxwell equations, $dF=0$ is a trivial identity in 1+1-dimensions, since it involves anti-symmetrizing three indices which can only take two different values. The other set of Maxwell equations, $d\,  ^{\star}F={^{\star}j}$, lead to $\epsilon^{ab}\partial_b E=-j^a$. These equations  imply that the electric field cannot vary out of the support of the sources $j$, neither in space nor in time. Hence, in 1+1-dimensions there are neither magnetic fields nor electric waves. 

Let us consider  a fixed, time dependent electric field that is uniform in space.  This can be generated by a time-dependent  current of the form $j^a=(0, j(t))$. We further assume that the electric field  is different from zero only  during a finite interval,  $E(t)\neq 0$ for $t_{\rm in}<t<t_{\rm out}$.  Despite the fact that $dF=0$ is an identity and there is no magnetic field, it is still useful to introduce a vector potential, $F=dA$, in terms of which the electric field reads ${E=\partial_t A_{\theta}-\partial_{\theta} A_t}$. A gauge transformation changes $A_{a}\to A_{a}+\partial_{a}\alpha$, with $\alpha$ a continuous function in the spacetime manifold. We can always use this freedom to make $A_t=0$, so that, under this gauge choice, $E=\partial_t A_{\theta}$.  
 
In 1+1-dimension, the expression \eqref{emanomaly} for the chiral anomaly is replaced by \cite{J85}.
\be \label{anomaly2d2}
\left\langle \nabla_{a} j_A^{a} \right\rangle=\frac{q\hbar }{2 \pi} \epsilon^{ab} F_{ab}\, . 
\ee
The right hand side can be written as $\frac{q\hbar }{2 \pi} \nabla_{a}j^{a}_{\rm CS}$, where $j^{a}_{\rm CS}=2\, \epsilon^{ab} A_{b}$. Although this vector is not gauge invariant, its divergence, as well as the scalar $Q_{CS}(t)\equiv  \int_t dx j^{0}_{CS}(t,x)$, are both gauge invariant.\footnote{A more rigorous derivation of this vector gives $j^{a}_{\rm CS}=2\, \epsilon^{ab} (A_{b}-A_b^0)$ with $\nabla_{[a} A_{b]}^0=0$, which includes an auxiliary potential $A_b^0$ that makes  $j^{a}_{\rm CS}$ gauge invariant. Since this extra term does not affect physical quantities, we can simply take $A_b^0=0$, as customary in the literature.  }
Following the argument described in the introduction, equation  \eqref{anomaly2d2} implies that the change of the chiral charge $\langle Q_A\rangle (t)= \int_{0}^L d\theta \langle j^{0}_A\rangle (t,\theta)$ from $t_{\rm in}$ to $t_{\rm out}$ can be written, for any quantum state, as
\be \label{DeltaQA2d} \langle Q_A\rangle (t_{\rm out}) - \langle Q_A\rangle (t_{\rm in})=\frac{q\hbar }{2 \pi} \,\left[ Q_{CS} (t_{\rm out}) -  Q_{CS} (t_{\rm in}) \right] \, , \ee
where $Q_{\rm CS}\equiv \int_{0}^L d\theta\, j^{0}_{\rm CS}=2\int_{\mathbb S^1}A_{a}d\ell^{a}$. As mentioned above, this quantity is manifestly gauge invariant. Recall also that this scalar is purely electric, i.e., it does not know anything abut the Dirac field. From this, we have
\be   Q_{CS} (t_{\rm out}) -  Q_{CS} (t_{\rm in})= 2 \int_{0}^L d\theta\, \big(A_{\theta}(t_{\rm out})-A_{\theta}(t_{\rm in})\big)= 2\, L\, \int_{t_{\rm in}}^{t_{\rm out}} dt \, E(t)\, , \ee
where in the last equality we have used that $E=\partial_t A_{\theta}$, that the electric  field is homogeneous, and that $L$ is the length of the spatial sections. 
Hence, Eqn.~\eqref{DeltaQA2d} tells us that the anomalous non-conservation of the chiral charge $\langle Q_A\rangle$ is dictated by the value of the time integral of the electric field
\be  \label{FDeltaQA2d}
 \langle Q_A\rangle (t_{\rm out}) - \langle Q_A\rangle (t_{\rm in})=\frac{q\hbar }{ \pi} \, L\, \int_{t_{\rm in}}^{t_{\rm out}} dt E(t)\, . \ee
 This result shows that the vacuum expectation value  $\langle Q_A\rangle $ ``keeps memory'' of the past history of the electric field. In particular, the effect of switching on an electric field  for some period of time $t_{\rm in}<t<t_{\rm out}$ can leave a residual, permanent helicity contribution on the vacuum state of the quantum field (quantified by the  value of $\langle Q_A\rangle$ at late times), {\it even} after switching off completely the external field. One can think about this residual helicity as the way the quantum field retains information about the past influence of the electric background.
 
 The integral on the RHS above also features in the  memory effect found for classical particles \cite{BG13}. A test charged particle in our background would  suffer a permanent  change in its velocity  after the passage of this electromagnetic pulse if and only if $\int_{t_{\rm in}}^{t_{\rm out}} dt E(t)\neq 0$. From this point of view,  the permanent change in the vacuum expectation value  $\langle Q_A\rangle$ found above can be thought of as another manifestation of the electromagnetic memory effect, but now on quantum fields.

It may be surprising that, despite the fact that the external  electric field vanishes,  the quantum system  does not return to its original configuration.  To  better understand this  effect,  one has to resort to the electromagnetic potential,  which makes manifest that  the memory actually originates   from  transitions between inequivalent  vacua of the electric field (hence, the background does not really return to the same exact configuration either). To see this, recall that our electric background evolves from a vacuum  configuration, $E(t_{\rm in}) = 0$, to another vacuum configuration $E(t_{\rm out}) = 0$. Classically, the two electric vacuum states are equivalent, but quantum mechanically they may not be.\footnote{The most prominent example of this is the Ahranov-Bohm effect \cite{AB59}.}.  In particular, note that the change in potential from $t_{\rm in}$ to $t_{\rm out}$, $A_{a}(t_{\rm out})-A_{a}(t_{\rm in})$, is non-trivial. This can be seen from the fact that the loop integral $\int_{\mathbb{S}^1} d\ell^{a} \, (A_{a}(t_{\rm out})-A_{a}(t_{\rm in}))$, which is gauge invariant,  is different from zero if and only if $\int_{t_{\rm in}}^{t_{\rm out}}dt  \, E(t)\neq 0$. {But since the electric field vanishes at early and late times, $A_{a}(t_{\rm out})$ can only differ from the initial potential $A_{a}(t_{\rm in})$ by a residual gauge transformation,  left by the dynamical evolution of the electric field.}
A straightforward calculation shows that this gauge transformation is given by $A_{a}(t_{\rm out})=A_{a}(t_{\rm in})+\partial_{a}\alpha$ with  $\alpha(\theta)=\theta\,  \int_{t_{\rm in}}^{t_{\rm out}} dt E(t)+\alpha_0$, for some constant $\alpha_0$. However, this is not an ordinary gauge transformation because $\alpha(\theta)$ is not a continuous function on $\mathbb{S}^1$ (because ${\alpha(L)\neq \alpha(0)}$). Instead, it belongs to the family of so-called ``large'' gauge transformations \cite{J85}, which carry physical implications, and which   can be used to label inequivalent notions of vacuum states of the quantum electromagnetic theory.\footnote{More precisely, when the potential $A_a$ is viewed as a gauge connection on a $U(1)$ principle bundle over $\mathbb R\times \mathbb S^1$, we can speak of infinitesimal gauge transformations, as well as of  global or finite gauge transformations. In the temporal gauge fixing $A_t=0$, a finite gauge transformation, $A_a \to A_a - i g^{-1}\nabla_a g$,  is determined by a continuous map $g: \mathbb S^1 \to \mathbb S^1$, given in a local coordinate system by $g(\theta)=e^{i\alpha(\theta)}$. Continuous maps on $\mathbb S^1$ can be divided in different (homotopy) classes, where two elements of the same class can be deformed continuously  into each other.  The classification of continuous maps is determined by the first homotopy group,  $\Pi(\mathbb S^1)\simeq \mathbb Z$, which shows that each class of gauge functions is labeled by an integer. This is easy to infer from the requirement that $g$ is  continuous, because it demands $\alpha(L)-\alpha(0)=2\pi n$, for $n\in \mathbb Z$. Two gauge functions $g$, $g'$ belonging to different classes cannot be deformed continuously into each other. An ordinary gauge transformation is a gauge transformation  $g$ that belongs to the trivial class or $n=0$, while gauge functions with $n\neq 0$ lead to ``large'' gauge transformations \cite{J85}.} 

In summary,  the passage of an electric pulse with $\int_{t_{\rm in}}^{t_{\rm out}} dt E(t)\neq 0$ induces a large gauge transformation in  the vector potential, which produces a memory effect, not only on classical charged particles, but also on the states of quantum fermion fields. As will be discussed below, in 3+1 dimensions there is another contribution to the chiral anomaly coming from the radiative content of the electromagnetic field; such contribution does not arise in 1+1 due to the absence of electromagnetic radiation.

The permanent change in the vacuum expectation value of the chiral charge $\langle Q_A\rangle$, described on the LHS of  (\ref{FDeltaQA2d}), can also be understood in terms of the standard notion of electromagnetic memory for particles. Heuristically, virtual charged particles populating the quantum vacuum would suffer  a permanent change in its velocity after switching on this electric pulse, provided the integral $\int_{t_{\rm in}}^{t_{\rm out}} dt E(t)$ does not vanish. In 1+1 dimensions these charges can only propagate in two directions, left or right. Therefore, positive charges suffer a ``kick'' in the direction of the electric field, while negative charges are kicked in the opposite direction. Both particles in the pair have the same helicity.  If the kick is strong enough, it will turn virtual charges into physical excitations out of the quantum vacuum.  This results in a net creation of helicity, which explains the permanent change of the quantum state or of the chiral charge $\langle Q_A\rangle$.

This heuristic picture can be made rigorous through a calculation of particle-pair creation using Bogoliubov coef- ficients. We finish this section with a brief allusion to this. If the electric field is non-zero only during a finite interval,  $t_{\rm in}<t<t_{\rm out}$, we can define natural ``in'' and ``out'' notions of vacua and particles. The question of interest is: if the field is prepared in the ``in'' vacuum before $t_{\rm in}$, and evolved until  a time after $t_{\rm out}$, what is the number of ``out'' quanta in the final state? 

This question can be answered without much difficulty in the case in which the electric field is uniform at all times (Appendix \ref{particlecreation2d} contains a detailed derivation of this and of the general case of an electric field that varies both in space and time).  As before, let us work in a gauge in which the vector potential is purely spatial, $A_t=0$. Without loss of generality, we can also consider $A_{\theta}(t)=0$ for $t<t_{\rm in}$. Let $A_0=\int_{t_{\rm in}}^{t_{\rm out}} \, E(t)\,  dt $ denote then the value of $A_{\theta}$ at late times, after $t_{\rm out}$. In short,  a non-zero value of $A_0$ induces a permanent frequency shift between the  ``in'' and ``out'' basis of solutions  of the field equations, which define the ``in'' and ``out'' vacua, respectively. Namely, for modes with spatial dependence $e^{i k \theta}$, with $k\in (2\pi/L)\, n$ and $n\in \mathbb{Z}$, the ``in'' modes oscillate with frequency $\omega_{\rm in}=k$, while ``out'' modes oscillate with frequency $\omega_{\rm out}=k+q\, A_0$. Because $k$ is discretized (due to the compactness of the spatial sections) there is a finite number of modes within the frequency interval $(0,q\, |A_0|)$. Namely, there are $\big[ q\, |A_0| \, \frac{L}{2\pi} \big]$ modes within this interval, where  the square brackets denote integer part.

This shift in frequency automatically implies that the evolution creates a number  $\big[ q\, |A_0| \, \frac{L}{2\pi} \big]$ of fermion-anti fermion pairs. Because of  linear momentum conservation (note that the background is homogeneous), anti-fermions (positive charges) move in the direction of $E(t)$, while fermions (negative charges) move in the opposite direction. However, helicity is not conserved in this process. Fermions moving to the right (left), and anti-fermions moving to the left (right), both have negative (positive) helicity. As a result, both members of the pair have positive $\hbar$     (negative $-\hbar$) helicity if $A_0>0$ ($A_0<0$). The total helicity carried by  the excited pairs is  $2\hbar \, \big[ q\, A_0 \, \frac{L}{2\pi} \big]=2\hbar \, \big[ q\,  \int_{t_{\rm in}}^{t_{\rm out}} \, E(t)\,  dt  \, \frac{L}{2\pi} \big]$. This agrees, except for the non-integer part, with the prediction for $ \langle Q_A\rangle (t_{\rm out}) - \langle Q_A\rangle (t_{\rm in})$ given  in Eqn.~\eqref{FDeltaQA2d}. The difference is due to the ``vacuum polarization'', i.e.\, the  helicity leftover in the ``out'' vacuum, which did not reach the threshold to excite another pair.

Although, for the sake of pedagogy, in this section we have restricted  to  uniform, background electric fields, all the arguments generalize to arbitrary functions $E(t,\theta)$. Appendix (\ref{particlecreation2d}) contains information about this generalization and further details which have been omitted in this section. \\

\section{Asymptotic structure of the electromagnetic field and infrared charges: a brief review}  \label{sec3}

The rest of this paper will focus on the chiral anomaly in asymptotically flat spacetimes in 3+1 dimensions. The presence of electromagnetic radiation, not present in 1+1 dimensions, makes it more convenient to use past and future null infinity for the initial and final Cauchy hypersurface of zero-rest mass fields. This section contains a brief summary of tools concerning the asymptotic structure of the electromagnetic field  at null infinity and infrared charges. These tools are well-known \cite{G77, A87, A14, ACL18}, and the reader familiar with them can skip this section.

\subsection{Review on the asymptotic structure of the electromagnetic field}

The electromagnetic radiation generated by  charges and currents can be rigorously studied  within the framework of asymptotically flat spacetimes \cite{G77, A87, A14}. This framework makes use of the notions of conformally compactified spacetimes introduced by Penrose in the 1960s \cite{P63}. 

Let (${\mathbb R}^4$, $ \hat\eta_{ab}$) represent the physical, Minkowski spacetime, and let ($M$, $ \eta_{ab}$) denote  an,  extended  (unphysical) spacetime obtained from (${\mathbb R}^4$, $ \hat\eta_{ab}$)  by an ordinary conformal compactification, i.e. by the addition of ``points at infinity''.\footnote{We shall use the hat symbol for any tensor and quantity intrinsic to the physical spacetime.} More precisely,  the new metric is obtained from the physical one by a conformal transformation $ \eta_{ab}=\Omega^2(x)  \hat\eta_{ab}$, while the new manifold is constructed by  attaching smoothly a null boundary $\mathcal J$ to the physical manifold, $M= {\mathbb R}^4\cup \mathcal J$. 
Locally, $\mathcal J$   corresponds to the hypersurface $\Omega=0$, and has null normal $\eta^{ab}\nabla_b \Omega\neq 0$. From a physical viewpoint, the elements of $\mathcal J$  represent the ``points of (null) infinity'', i.e. the points that can be asymptotically reached by following radial, null geodesics  in the physical spacetime. The boundary $\mathcal J$ is made of two portions, past ($\mathcal J^-$) and future  ($\mathcal J^+$)  null  infinity. In the following, we will focus on $\mathcal J^+$. The construction is similar for $\mathcal J^-$.  

For example, in a Bondi-Sachs coordinate system $\{u,r,\theta,\phi\}$, where $u=t-r$ is the standard retarded time, the Minkowski metric reads $d\hat s^2=-du^2+2dudr+r^2d\omega^2$ and one uses $\Omega=1/r$ to obtain $d s^2=-\Omega^2du^2+2dud\Omega+d\omega^2$ after the conformal transformation mentioned above. The restriction of this line element to the $\Omega=0$ hypersurface gives a  well-defined (although degenerate) metric. The limit $r\to \infty$  keeping $u,\theta,\phi$ constant follows the geodesics of outgoing radiation propagating to future null infinity, getting to $\Omega=0$ in finite time as measured by the unphysical metric. The extended manifold is obtained then by including  all these limiting points $\{u,\Omega=0,\theta,\phi\}$ to the original manifold, and future null infinity is described then by the submanifold $\mathbb R\times \mathbb S^2$.

This framework makes it possible to study the behavior of  the electromagnetic field in a neighborhood of infinity (which in this case  is simply a boundary of the  spacetime manifold)
using  standard techniques in differential geometry. To see this, let us first note that the electromagnetic field tensors are conformal invariant, $\hat F_{ab}=F_{ab}$, $\hat A_a=A_a$. These tensors  are well-defined in the entire extended spacetime, including at the boundary $\mathcal J^+$. The electromagnetic field $F_{ab}$ has six independent components. In a Newman-Penrose basis $\{n^a, \ell^a,m^a,\bar m^a\}$ \cite{NP62}, where typically one takes $\ell_a=-\nabla_a u$ as the  vector tangent to outgoing null geodesics, the 6 electric and magnetic  components of $F_{ab}$ can be captured in the following 3 complex scalars
\bea
\label{phi2} \Phi_2 & := & F_{ab}n^a \bar m^b\, ,\\
\label{phi1}  \Phi_1 & := & \frac{1}{2} \left[ F_{ab}n^a \ell^b+F_{ab} m^a \bar m^b \right]  \, ,\\
\label{phi0}  \Phi_0 & := & F_{ab}m^a \ell^b  \, .
\eea
 If we assume smooth fields, the Peeling theorem guarantees that these scalars  admit the following Taylor expansion in $\Omega$  in a neighborhood of future null infinity \cite{stewart}:
\bea
\label{phi20}\Phi_2(u,\Omega,\theta,\phi) & = & \Phi_2^{0}(u,\theta,\phi)+ \Omega \Phi_2^{1}(u,\theta,\phi) +\dots\, ,\\
\label{phi10}\Phi_1(u,\Omega,\theta,\phi) & = & \Phi_1^{0}(u,\theta,\phi)+ \Omega \Phi_1^{1}(u,\theta,\phi) +\dots\, ,\\
\label{phi00}\Phi_0(u,\Omega,\theta,\phi) & = & \Phi_0^{0}(u,\theta,\phi)+ \Omega \Phi_0^{1}(u,\theta,\phi) +\dots\, ,
\eea
where we denote $\Phi_2^{0}(u,\theta,\phi)\equiv  \Phi_2(u,\Omega=0,\theta,\phi)$ and similarly for $ \Phi_1^{0}(u,\theta,\phi)$ and $\Phi_0^{0}(u,\theta,\phi)$. These fields encode all the information about the electromagnetic field at $\mathcal J^+$.  They are, however, not independent. Using Maxwell's equations, one finds
 \bea
 \partial_u \Phi_1^0 = \eth \Phi_2^0\, ,  \label{maxwell1}\\
 \partial_u \Phi_0^0 = \eth \Phi_1^0\, ,\label{maxwell2}
 \eea
  where  $\partial_u f=n^a \nabla_a f$ for any function $f$ (this is a consequence of the Newman-Penrose normalization $n^a\ell_a=-1$), and where $\eth$ is a spin-weighted derivative operator \cite{stewart}, defined by $V_1^b V_2^c \dots m^a\nabla_aT_{bc\dots}=\eth (V_1^b V_2^c \dots T_{bc\dots})$, for  arbitrary tensors $V_i^a$ and  $T_{bc\dots}$.
 These equations determine the evolution of the scalars $\Phi_0^0$ and $\Phi_1^0$ along the retarded time $u$ in $\mathcal J^+$,  upon giving initial conditions, and also some input for $\Phi_2^0$. In contrast, the dynamics of $\Phi_2^0$ along $u$ is not determined by  Maxwell equations. This scalar serves as the free data for a characteristic value formulation of Maxwell theory at $\mathcal J^+$.  
 
By switching back to the original physical spacetime, with the appropriate conformal rescaling of the Newman-Penrose vectors, one can see that $\hat\Phi_2\sim O(r^{-1})$, and $\hat\Phi_1\sim O(r^{-2})$. From this, one identifies the scalar $\Phi_2^0$ as describing the two radiative degrees of freedom of the electromagnetic field, while $\Phi_1^0$ represents the Coulombic part of the field. In fact, the total energy flux radiated  to $\mathcal J^+$ is given by $F=\int dud\mathbb S^2 \, T_{ab}n^an^b =\int dud\mathbb S^2 |\Phi_2^0|^2$, where $T_{ab}$ is the energy-momentum tensor, and therefore is entirely determined from $\Phi_2^0$.
Similarly, the electric charge of sources in the bulk  can be determined from $\mathcal J^+$ using  Gauss' Law as $Q=\frac{1}{2\pi}\int d\mathbb S^2 {\rm Re} \,\Phi_1^0(u,\theta,\phi)$, and is completely determined from $\Phi_1^0$. Using Maxwell equations in $\mathcal J^+$, it is straightforward to check that $\partial_u Q=0$, reflecting the conservation of the electric charge.  Note also  that the requirement of finite energy flux at $\mathcal J^+$ , $F<\infty$, requires $\Phi_2^0(u,\theta,\phi)\to 0$ as $u\to \pm \infty$.

In terms of an electromagnetic potential, one introduces the scalars $A_2:=A_a \bar m^a$, $A_1=A_a n^a$, $A_0 = A_a \ell^a$. The fall-off conditions of the potential for large $r$ is not given beforehand from the theory and requires some input. Physical considerations require that these components admit an asymptotic series with leading behavior $O(r^{-1})$ \cite{AB17a, AB17b}. Using $F=dA$, one can obtain the following formulas, valid at future null infinity $\mathcal J^+$:
\bea
\Phi_2^0 & = & \dot A_2^0 - \bar \eth A_1^0\label{phi22} \, ,\\
{\rm Im} \,\Phi_1^0 & = & {\rm Im}\, \eth A_2^0 \, , \label{imphi1}
\eea
where dot denotes derivative with respect to retarded time $u$.
Furthermore, by integrating Maxwell equations at $\mathcal J^+$, one can further obtain:
\bea
{\rm Re}\, \Phi_1^0 & = & {\rm Re}\, \eth A_2^0 - \int^u_{u_0}du' \eth \bar\eth A_1^0 + G(u_0,\theta,\phi)\, ,\label{rephi1}
\eea
where $G(u_0,\theta,\phi)$ arises as an integration factor. 
From \eqref{phi2} we see that the two electromagnetic radiative degrees of freedom are distributed between $A_2^0$ and $A_1^0$. But this is 3 real-valued scalars, so there is, as expected, some gauge redundancy in the description. A useful gauge fixing is $A_1^0\equiv A_a n^a=0$. With this gauge choice, the real and imaginary parts of $A_2^0$ represent the two radiative degrees of freedom,  electric and magnetic respectively. The Coulombic aspects of the field are all encoded in $G(u_0,\theta,\phi)$, in particular $Q=\frac{1}{2\pi}\int_{\mathbb S^2} G(u_0,\theta,\phi)$.

\subsection{Electromagnetic soft charges and the memory effect}

The phenomenon of memory effect is well-known in Maxwell's theory \cite{BG13}. 
The most prominent example is  a charged point-like particle of initial velocity $\vec v_1$ that suffers a ``kick'',  and changes  its direction of propagation to   $\vec v_2$ after the passage of an electromagnetic pulse. This is an example of electric-type memory. In the intermediate process,  the charged particle emits radiation by Bremsstrahlung; the properties of the emitted radiation  carry information about this effect. 

At future null infinity, the phenomenon of electromagnetic memory is encoded in the following quantities:
\bea  \label{softcharges}
q_\alpha=\int d \mathbb S^2 \alpha(\theta, \phi)\left(\Phi_1^0(\infty, \theta, \phi)-\Phi_1^0(-\infty, \theta, \phi)\right)\, , \eea
where  $\alpha$ is a smooth real function  on the sphere $\mathbb S^2$.  The complex numbers $q_\alpha$ are called soft charges, and they measure    permanent changes  in the multipolar structure of the Coulombic part of the electromagnetic field after some process. From (\ref{maxwell1}),  one infers that $q_\alpha\neq 0$ only if $\Phi_2(u,\theta,\phi)\neq 0$, i.e. only, if there is a flux of electromagnetic radiation reaching infinity. When this happens, one says that the electromagnetic field keeps  memory on the radiation flux emitted to infinity in the past.

The relation of the charges  $q_\alpha$ and the radiation reaching $\mathcal J^+$ can be explicitly shown by   using Maxwell equations to replace $\Phi_1^0$ by $\Phi_2^0$ in \eqref{softcharges}:
\bea
\label{softchargeem} q_{\alpha}  =\int d \mathbb S^2 d u \alpha(\theta, \phi) \dot{\Phi}_1^0(u, \theta, \phi)=\int d \mathbb S^2 d u \alpha \eth \Phi_2^0(u, \theta, \phi)=-\int d \mathbb S^2 du  \eth\alpha(\theta, \phi)  \Phi_2^0(u, \theta, \phi) \, .
\eea
Expanding  in spin-weighted spherical harmonics, this can be further simplified as 
\bea
q_{\alpha} = \sum_{\ell m} (-1)^m\sqrt{\ell(\ell+1)}\alpha_{\ell -m}\int_{-\infty}^{+\infty}du\Phi^0_{2\ell m}(u)\equiv \sum_{\ell m} \tilde\alpha_{\ell m} q_{\ell m} \, .
\eea
The problem is reduced to study the basis $q_{\ell m}=\int_{-\infty}^{+\infty}du\Phi^0_{2\ell m}(u)$ of charges. Note that, for $\ell=0$, i.e. when $\alpha(\theta,\phi)=$ const, the soft charge is identically  zero. This is consistent with the fact that the monopole of the electromagnetic field (the electric charge) is conserved and cannot be radiated away. In contrast, dipolar and higher order structure ($\ell\geq 1$) can be radiated away. That phenomena is encoded  in $q_{\ell m}\neq 0$ for $\ell>0$. 

As argued in the previous subsection, finiteness of energy fluxes require that $\Phi_{2\ell m}(u)$ belongs to $L^2(\mathbb R, \mathbb C)$, which implies that   it  admits a Fourier transform on $\mathcal J^+$:
\be
\tilde{\Phi}^0_{2\ell m}(\omega)=\int_{-\infty}^{+\infty} d u\,  \Phi^0_{2 \ell m}(u) \, e^{-i \omega u} \, .  \ee
This automatically implies that the charges $q_{\ell m} $ are simply the ``zero mode'' of $\tilde{\Phi}_{2\ell m}^0$
\be  q_{\ell m}=\tilde{\Phi}^0_{2\ell m}(0) \, . \label{zeromode}
\ee
Therefore, only the zero-frequency modes of the emitted electromagnetic radiation   leave a memory on the multipolar structure of the field. This is the reason why these charges are  called ``soft'', as they are associated with ``soft photons'' \cite{W64, S17}.

These charges are intrinsically associated to asymptotic symmetries of Maxwell theory.
 One way of looking into this is by considering the  phase space $\Gamma$ of the electromagnetic degrees of freedom at $\mathcal J^+$. This phase space is made of pairs of canonically conjugate fields\footnote{In the Newman-Penrose basis introduced above, notice that the tangent space at each point of future null infinity is spanned by the three vectors  $\{n^a, m^a,\bar m^a\}$, while the cotangent space is spanned by the three co-vectors  $\{\ell_a,m_a,\bar m_a\}$.} $(A_a, \mathcal E^b)$, where $A_a=-A_1^0 \ell_a +A_2^0 m_a+\bar A_2^0 \bar m_a$ and $\mathcal E^a=-2{\rm Re}\, \Phi_1^0 n^a +\Phi_2^0 m^a+\bar \Phi_2^0 \bar m^a$\footnote{No gauge condition on $A_a$ has been imposed at this level.}.  $\Gamma$ can be endowed with the structure of an infinite-dimensional Banach manifold.
The usual symplectic structure for Maxwell theory can be written on future null infinity as 
\bea
\Omega(\, (A_a^{(1)},\mathcal E^{(1)a}),(A_a^{(2)},\mathcal E^{(2)a})\, )=\int_{\mathbb R\times\mathbb S^2} du \, d\mathbb S^2 \left(A_a^{(1)}\mathcal E^{(2)a} - A_a^{(2)}\mathcal E^{(1)a} \right)\, .
\eea
Together with  suitable fall-off conditions at $u\to\pm\infty$ required  to make this integral well-defined, the pair $(\Gamma,\Omega)$ defines the phase space for the  radiative degrees of freedom of Maxwell theory.

Now, consider the (restricted) family of  gauge transformations $A_a\to A_a+D_a \alpha$, with $\alpha=\alpha(\theta,\phi)$. This transformation is generated in phase space by the quantity
\bea
Q_{\alpha}\equiv \frac{1}{2}\Omega(\, (A_a,\mathcal E^{a}),(D_a\alpha,0)\, ) & = & - \frac{1}{2}\int_{\mathbb R\times\mathbb S^2} du \, d\mathbb S^2  (D_a\alpha) \mathcal E^{a} = - \frac{1}{2}\int_{\mathbb R\times\mathbb S^2} du \, d\mathbb S^2  D_a(\alpha \mathcal E^{a})  = \int_{\mathbb R\times\mathbb S^2} du \, d\mathbb S^2 n^a D_a(\alpha {\rm Re}\, \Phi_1^0) \nonumber\, ,
\eea
where in the second equality we have used   Maxwell equations  (\ref{maxwell1}) to write  $D_a \mathcal E^{a}\propto  {\rm Re}\, \dot\Phi_1^0 - {\rm Re}\,\eth \Phi_2^0=0$.  Recalling that $\partial_u=n^aD_a$, the RHS of this equation happens to be equal to the real part of the soft charges defined in (\ref{softcharges}). (A similar analysis using a  ``dual'' potential $Z_a$ produces the imaginary part of the soft charges).

This observation tells us that  soft charges $q_{\alpha}$ can be identified with the generators of gauge transformations in the radiative phase space. Since soft charges can be different from zero, one concludes that transformations $A_a\to A_a+D_a \alpha$ in $\mathcal J^+$ are  actual  symmetries of our phase space $(\Gamma,\Omega)$, rather than  mere  gauge transformations. Therefore, they have  physical significance (which is, precisely, the electromagnetic memory). From the viewpoint of the bulk, these are gauge transformations that do not vanish at infinity. To distinguish them from ordinary gauge transformations, they are called ``large'' gauge transformations. The set of all large gauge transformations constitutes the infinite-dimensional, asymptotic symmetry group of Maxwell theory.

\section{The chiral anomaly induced by an electromagnetic background} \label{sec4}

This section contains the main results of  this article.  We consider a quantum, massless Dirac  field  propagating in Minkowski spacetime in 3+1 dimensions with metric $\eta_{ab}$, coupled to an electromagnetic field $F_{ab}$. The spin 1/2 field is treated as a test field, i.e. we neglect its back-reaction on the electromagnetic and spacetime backgrounds.   This external electromagnetic field  is assumed to be generated by some distribution of electric charges and currents,   that are smooth and confined in space, but otherwise arbitrary. To keep the parallelism with  the 1+1-dimensional chiral anomaly discussed in the previous section,  the sources will be ``switched on'' only for a finite amount of time, in the sense that they become stationary at sufficiently late and early times. All possible electromagnetic waves are radiated  during a finite period of time. 
 
As discussed  above, the electromagnetic field  can induce a change in the helicity of the fermionic field due to the chiral anomaly.  Our starting point is expression \eqref{integral} for the change of the chiral charge of the quantum field 
\be 
\label{deltaQA2} \left< Q_A\right>_{\Sigma_{\rm out}}- \left< Q_A \right>_{\Sigma_{\rm in}}= \int_R  d^4x\sqrt{-\eta}\, \left< \nabla_{a}j^{a}_A\right>=-\frac{\hbar q^2}{8\pi^2}\int_R d^4x \sqrt{-\eta} F_{ab}\, ^{\star}F^{ab} \, . 
\ee 
If we integrate over  the entire spacetime manifold, $R=M\simeq \mathbb R^4$, then $\Sigma_{\rm in}$ and $\Sigma_{\rm out}$ correspond to past and future null infinity, respectively. This choice makes it possible to use the machinery summarized in the previous section to disentangle the properties of the electromagnetic field that can make the RHS  different from zero. This  problem was worked out in  \cite{dR21}, where it was found that, assuming no incoming electromagnetic radiation from past null infinity,  the  RHS of \eqref{deltaQA2} can be written in terms of  boundary data on future null infinity as
\bea
 \left< Q_A\right>_{\mathcal J^{+}}- \left< Q_A \right>_{\mathcal J^{-}}= \frac{\hbar q^2}{4 \pi^2} \int_{-\infty}^{\infty} d u \int d \mathbb S^2 {\rm Im} \left\{\left(A_2^0-\eth \alpha_0\right) \bar\Phi_2^0\right\} \label{electromagnetism} \, .
\eea
Here, $\alpha_0$ is a smooth real-valued function on the sphere and  $\eth \alpha_0$ is a pure gauge potential (i.e. it produces no electromagnetic field, $\Phi_2^0=0$). This expression was derived in the gauge $A_1^0=0$. Notice that a non-zero value is obtained in the integral (\ref{electromagnetism}) because of the  weak decay behavior of the radiative  solutions of Maxwell equations in a neighboorhood of future null infinity: $A_2\sim 1/r$, $\Phi_2\sim 1/r$ (recall the discussion of page 6). These two radial factors compensate the $r^2$ factor in the integral measure. \footnote{There are no contributions to (\ref{deltaQA2}) from spatial infinity, nor from future or past timelike infinities, because the product of the electromagnetic field and its potential decays too fast (as $O(r^{-3})$) in those directions, see \cite{dR21} for details.}.

To analyze the physical interpretation of the RHS of the previous equation it is convenient to work with a compactified  retarded coordinate $u$. We will consider $u\in [-L/2, L/2]$ and let $L\to \infty$ at the end of the calculation. As explained above, if $\Phi_2^0$ is the electromagnetic radiation field, the requirement that the energy flux across $\mathcal J^+$ is finite implies $\Phi_2^0(\cdot, \theta,\phi)\in L^2(\mathbb R)$ for all $(\theta,\phi)\in \mathbb S^2$, and in particular $\Phi_2^0\to 0$ as $u\to \pm \infty$. Therefore,  we will consider functions $\Phi_2^0(\cdot, \theta,\phi)\in L^2(  (-L/2, L/2) )$, for all $(\theta,\phi)\in \mathbb S^2$, with boundary conditions given by $\Phi_2^0(\pm\frac{L}{2},\theta,\phi)=0$. This will guarantee that $\Phi_2^0\to 0$ as $u\to \pm \infty$ at the end of the calculation. Since the functions $\Phi_2^0(\cdot,\theta,\phi)$ happen to be periodic with period $L$, an orthonormal basis for $L^2(  (-L/2, L/2) )$ is given by $\{\frac{e^{-i\omega_n L/2}}{\sqrt{L}}e^{-i\omega_n u}\}_{n \in \mathbb Z}$, where $\omega_n=\frac{2\pi}{L}n$, so one can expand in Fourier series:
\bea
\Phi_2^0\left(u_, \theta, \phi\right)=\sum_{n=-\infty}^{+\infty} \tilde \Phi_2^0\left(\omega_n, \theta, \phi\right) e^{-i \omega_n \frac{L}{2}} \frac{e^{-i \omega_nu} }{\sqrt{L}} \, .
\eea
The inverse Fourier series is
\bea
\tilde \Phi_2^0\left(\omega_n, \theta, \phi\right)= e^{i \omega_n \frac{L}{2}} \frac{1}{\sqrt{L}} \int_{-L / 2}^{L / 2} \Phi_2^0(u, \theta, \phi) e^{+i \omega_n u} d u\, .
\eea
The basis modes are orthonormal with respect to the $L^2$ norm:
\bea
\int_{-L / 2}^{L / 2}du \frac{1}{\sqrt{L}} e^{-i \omega_n u}(-1)^n \frac{1}{\sqrt{L}} e^{+i \omega_{n^{\prime} } u }(-1)^n=\delta_{n n^{\prime}} \label{ortho} \, .
\eea
The continuous limit will be recovered using the formula $\lim _{L \rightarrow \infty} \frac{1}{L} \sum_{n \in \mathbb Z} f\left(\frac{n}{L}\right)=\frac{1}{2 \pi} \int_{-\infty}^{+\infty} d w f(w)$. 

In order to disentangle the potential contribution of  IR charges to equation (\ref{electromagnetism}) we will make an explicit distinction between the zero frequency mode $\tilde \Phi_2^0(0,\theta,\phi)\neq 0$ and the rest of modes. Let us then write the field and potential as
\bea
\Phi_2^0\left(u_, \theta, \phi\right) & = & \sum_{n\neq 0} \tilde \Phi_2^0\left(\omega_n, \theta, \phi\right) (-1)^n \frac{e^{-i \omega_nu} }{\sqrt{L}} + \frac{1}{\sqrt{L}} \Phi_2^0\left(0, \theta, \phi\right)\, , \label{equationphi2}\\
A_2^0\left(u, \theta, \beta\right) &  = &\frac{1}{\sqrt{L}} \sum_{n \neq 0} \tilde\Phi_2^0\left(u, \theta,\phi\right) \frac{e^{-i \omega_n u}-e^{-i \omega_n u_0}}{-i \omega_n}(-1)^n+\frac{u-u_0}{\sqrt{L}} \tilde{\Phi}_2\left(0, \theta, \phi\right)+A_2^0\left(u_0, \theta, \phi\right)\, , \label{equationA}
\eea
where the second line is derived using $\Phi_2^0=\dot A_2^0$, which is valid in the gauge $A_1^0=A_a n^a=0$. 
Next, we substitute this expansion in \eqref{electromagnetism} and keep track of the contribution of the zero mode. The calculation is tedious, and is written in detail in  Appendix \ref{appendixIR}. We focus here in the result and its physical meaning:
\bea
 \left< Q_A\right>_{\mathcal J^{+}}- \left< Q_A \right>_{\mathcal J^{-}}=\frac{\hbar q^2}{4 \pi^2} \left[ \frac{1}{2\pi} \int d \mathbb S^2 \int_{-\infty}^{\infty} d\omega \frac{\left|\tilde\Phi_2^0(\omega, \theta, \phi)-\tilde\Phi_2^0(0, \theta, \phi)\right|^2}{\omega} + {\rm Im}q_{\alpha} -{\rm Re} q_{\beta}\right] \, . \label{final2}
\eea
The first term in the RHS of this equation contains the contribution from electromagnetic radiation   with non-zero frequencies reaching  $\mathcal J^{+}$; the subtraction of $\tilde\Phi_2^0(0, \theta, \phi)$ removes the zero mode from the integral. This, in turn, makes the integrand finite in the limit $\omega \to 0$ (the integral is well-defined for $\omega\to \pm \infty$ by Plancherel Theorem). This term can be further expressed as \cite{dR21}:
\be   \int_{-\infty}^{\infty} d\omega \frac{\left|\tilde\Phi_2^0(\omega, \theta, \phi)-\tilde\Phi_2^0(0, \theta, \phi)\right|^2}{\omega}=  \int_{0}^{\infty} \frac{d\omega}{\omega}\,  \left( \left|{\tilde \Phi}_{\rm R}(\omega, \theta, \phi)\right|^2-\left|{\tilde\Phi}_{\rm L}(\omega, \theta, \phi)\right|^2\right)\,  , \label{hscri}\ee
where $ \Phi_{\rm R}(\omega, \theta, \phi):=\tilde\Phi_2^0(\omega, \theta, \phi)-\tilde\Phi_2^0(0, \theta, \phi)$ defined for $\omega>0$, and $ \bar\Phi_{\rm L}(\omega, \theta, \phi):=\tilde\Phi_2^0(-\omega, \theta, \phi)-\tilde\Phi_2^0(0, \theta, \phi)$ defined for $\omega<0$, describe  right- and left-handed circularly polarized radiation, respectively.   Expression \eqref{hscri} has a neat physical interpretation: it measures the net electromagnetic helicity radiated to $\mathcal J^{+}$. 

The second and third terms in \eqref{final2} come entirely from the zero mode of the electromagnetic field, and correspond to two infrared charges of magnetic and  electric type, ${\rm Im}q_{\alpha}$ and ${\rm Re} q_{\beta}$, respectively:
\bea {\rm Im}\, q_{\alpha}&=&   \int d \mathbb S^2 \alpha(\infty, \theta, \phi)\left( {\rm Im}\, \Phi_1^0(\infty, \theta, \phi)- {\rm Im}\,\Phi_1^0(-\infty, \theta, \phi)\right) \, , \label{magneticq} \\
  {\rm Re} \, q_{\beta}&=&   \int d \mathbb S^2 \beta(\infty, \theta, \phi)\left( {\rm Re}\,\Phi_1^0(\infty, \theta, \phi)- {\rm Re}\,\Phi_1^0(-\infty, \theta, \phi)\right) \,  . \label{electricq} 
 \eea
In these equations the real-valued functions $\alpha$ and $\beta$ are defined from the longitudinal and transverse part of the electromagnetic potential at future timelike infinity, as follows. In the gauge we are using,   in which $A_1^0=A_a n^a =0$, the 1-form $A_a^0$ lives on the cotangent space of  each cross section $\mathbb S^2$ of $\mathcal J^{+}$. Therefore, it can be expressed as the sum of a gradient and a curl: $A_a^0=D_a \alpha+\epsilon_a^{\hspace{0.15cm}b}D_b\beta$, where $D_a$ is the covariant derivative on $\mathbb S^2$. This equation defines $\alpha(u,\theta, \phi) $ and $\beta(u,\theta, \phi) $.  Using (\ref{imphi1})-(\ref{rephi1}), they can be solved from ${\rm Re} \,\Phi_1^0(u,\theta,\phi)=\Delta \alpha(u,\theta,\phi)+G(\theta,\phi)$ and ${\rm Im}\, \Phi_1^0(u,\theta,\phi)=\Delta \beta(u,\theta,\phi)$, where $\Delta$ denotes the 2-dimensional Laplacian.   Since the soft charges ${\rm Im}\, q_{\alpha}$ and ${\rm Re}\, q_{\beta}$ are specified from  functions $\alpha$ and $\beta$, which depend on the electromagnetic potential, these are {\it field-dependent} soft charges.   Since $\alpha$, $\beta$ originate from a gradient and a curl, respectively, the function $\alpha$ can be thought of as the electric  degree of freedom of the emitted waves, while $\beta$ is the magnetic one.

In summary,  the change  of the chiral charge of a quantum, massless, Dirac field  between  past to future null infinity, resulting from its coupling to an electromagnetic background, yields
\be   \label{main} \left< Q_A\right>_{\mathcal J^{+}}- \left< Q_A \right>_{\mathcal J^{-}}= \frac{\hbar q^2}{4 \pi^2} \left[\frac{1}{2\pi}  \int d \mathbb S^2\int_{0}^{\infty} \frac{d\omega}{\omega}\,  \left( \left|{\tilde \Phi}_{\rm R}(\omega, \theta, \phi)\right|^2-\left|{\tilde\Phi}_{\rm L}(\omega, \theta, \phi)\right|^2\right) + {\rm Im}q_{\alpha} -{\rm Re} q_{\beta} \right]\, . \ee
This is the main result of this paper. It  shows that the anomalous non-conservation of fermionic helicity receives two types of contributions from an external electromagnetic field. Namely, $\left< Q_A\right>$ can change in time if (1) a distribution of electric currents and charges in the bulk are able to radiate chiral electromagnetic waves, and (2) there is a change in the infrared sector of the external electromagnetic field, such that  the two soft charges (\ref{magneticq})-(\ref{electricq}) are different from zero. The presence of $\hbar$ emphasizes that this is a quantum effect with no classical analog; it originates from the chiral anomaly.

We finish this section with a few remarks.

{\bf Remark 1.} We have assumed no incoming radiation from ${\mathcal J^{-}}$. If the electromagnetic field is not trivial  at past null infinity, we just need to replace quantities at ${\mathcal J^{+}}$ above with differences between ${\mathcal J^{+}}$ and ${\mathcal J^{-}}$.

{\bf Remark 2.} The contribution from soft charges bears some similarity with the rationale behind the theory of instantons \cite{J77, S94}, in which quantum-mechanical transitions between ``topologically inequivalent'' vacuum states of the Hilbert space underlying a non-abelian gauge theory induces an anomaly. In the quantum theory of the electromagnetic field, for each non-trivial IR sector one has a representation of the canonical commutation relations which is unitarily inequivalent to the usual Fock representation. So, just like with the interpretation of the instantons, we can say here that tunneling transitions between the different IR vacuum states of the electromagnetic field induces the fermionic chiral anomaly \footnote{Notice that this was precisely the origin of the chiral anomaly in 1+1 dimensions discussed in section \ref{sec2}.}. In contrast, in this approach 
there is no need to work with Euclidean field equations. In fact, by working with solutions of the Lorentzian Maxwell equations we also get a radiative contribution, in addition to the contribution from soft charges. This radiative contribution is not predicted in the Euclidean case, where everything is ``instantaneous''.\\

\subsection{Examples}

 We discuss now  examples of electromagnetic sources that are able to trigger  the chiral anomaly obtained in (\ref{main}). A physical configuration of electric charges and currents that can radiate circularly polarized electromagnetic waves was described in \cite{dR21}, namely, an electric-magnetic oscillating dipole.  In this subsection we focus  on  examples that produce non-zero values of the  infrared charges (\ref{electricq}) and (\ref{magneticq})

Soft charges of electric-type are determined by the Coulombic contribution (i.e. $\sim 1/r^2$) of the radial component of the electric field:
\bea
\operatorname{Re} q_{{\ell m }}=\left.\oint d \mathbb S^2 \operatorname{Re} \Phi_1^0(u, \theta, \phi) Y_{\ell m}(\theta, \phi)\right|_{-\infty} ^{+\infty}, \quad \quad \operatorname{Re} \Phi_1^0(u, \theta, \phi)=\lim _{r \rightarrow \infty}\left[r^2 E_r\left(u,r, \theta, \phi\right)\right] \label{electricformula}\, .
\eea
The canonical example where these soft charges are not zero is a charged particle with some initial velocity that interacts with an external source (a nucleus, another charged particle, etc.) and changes its velocity \cite{BG13}. In this process, the charged particle emits Bremstrahlung, which is known to possess zero-frequency photons. We shall review this example here for completeness. 

The electromagnetic field generated by a moving charged particle can be obtained in closed form from the Lienard-Wiechart potentials \cite{Griffiths}:
\bea
\vec{E}(t, \vec{r}) & = & \left.\frac{q}{4 \pi\left(1-{\vec n_s \cdot \vec{v_s}}\right)^3}\left[\frac{\vec{n}_s-{\vec v_s}}{\left(1-v_s^2\right)\left\|\vec{r}-{\vec r_s}\right\|^2}+\frac{\vec n_s \times\left[\left({\vec n_s}-{\vec v_s}\right) \times \dot{\vec v}_s \right]}{\left\|\vec{r}-{\vec r_s}\right\|}\right]\right|_{t=t_r}\, , \label{electricLW}\\
\vec{B}(t, \vec{r}) &  = & {\vec v_s}\left(t_s\right) \times \vec{E}(t, \vec{r})\, , \label{magneticLW}
\eea
where $\vec r_s(t)$ is the location of the charge, $\vec v_s=\dot{\vec r}_s$ its velocity, $\vec n_s=\frac{\vec r-\vec r_s(t)}{||\vec r-\vec r_s(t)||}$,  and $t_r=t-||\vec r-\vec r_s(t_r)||$ is the retarded time (which is a function of $(t,\vec r)$). For a particle with constant velocity, $\dot{\vec v}_s=0$, and the electric field above can be rewritten as \cite{Griffiths}
\bea
\vec{E}(t, \vec{r})=\frac{q}{4 \pi} \frac{1-v^2}{\left[1-v^2+\left(\frac{(\vec{r}-\vec{v} t) \cdot \vec{v}}{\| \vec{r}-\vec{v}t \|}\right)^2\right]^{3 / 2}} \frac{\vec r-\vec{v} t}{\|\vec{r}-\vec{v} t\|^3}\, , \quad\left(\vec r_s(t)=\vec{v}_s \cdot t\right)\, .\label{constantv}
\eea
Changing to Bondi-Sachs coordinates $\{u,r,\theta,\phi\}$, and taking the limit $r\to \infty$ keeping $\{u,\theta,\phi\}$ constant, one  obtains the following expression for the radial component of the electric field
\bea
E^a(u,r,\theta,\phi)\nabla_a r = \frac{q}{4 \pi r^2} \frac{1-v^2}{\left(1-v^a \nabla_a r \right)^2} =  \frac{q}{4 \pi r^2} \frac{1-v^2}{\left(1-v \cos\theta \right)^2} \, ,
\eea
where in the last equality  we have chosen $\hat z$ in the direction of $\vec v_s$. From this expression and (\ref{electricformula}) one readily obtains
\bea
 \operatorname{Re} \Phi_1^0(u, \theta, \phi) =  \frac{q}{4 \pi } \frac{1-v^2}{\left(1-v \cos\theta \right)^2} \, . \label{ex1}
\eea
For a particle that always moves with the same constant velocity,  $\operatorname{Re} q_{{\ell m }}=0$ for all $\ell$. This is easy to see if we choose the reference system comoving with the particle, so that $\vec v_s=0$ and $ \operatorname{Re} \Phi_1^0(u, \theta, \phi) =  \frac{q}{4 \pi }$.
However, if the particle interacts with some external potential and changes its velocity, then we can no longer choose an inertial reference system attached to the particle at all times. While at early times we may have $ \operatorname{Re} \Phi_1^0(u\to-\infty, \theta, \phi) =  \frac{q}{4 \pi }$, at late times we will have $ \operatorname{Re} \Phi_1^0(u\to+\infty, \theta, \phi) = \frac{q}{4 \pi } \frac{1-v^2}{\left(1-v \cos\theta \right)^2}$. The electric soft charges can be now computed:
\bea
\operatorname{Re} q_{{\ell m }}=\frac{q}{4 \pi}\left(1-v^2\right) \oint_{i^+}d\mathbb S^2 \frac{Y_{\ell m}(\theta,\phi)}{(1-v \cos \theta)^2} - q \delta_{\ell 0} = \frac{q\left(1-v^2\right)}{2}\delta_{m0} \int_{-1}^1 d x \frac{P_{\ell}(x)}{(1-v x)^2} - q \delta_{\ell 0} \, .
\eea
In particular, $\operatorname{Re} q_{{0 0 }}=0$ due to electric charge conservation, as expected. By taking different values of $\ell$ one can check that this expression is indeed different from zero. In summary, soft charges of electric type can be generated by Lorentz boosting an electric charge.

 Soft charges of magnetic-type are determined by the Coulombic contribution (i.e. $\sim 1/r^2$) of the radial component of the magnetic field:
\bea
\operatorname{Im} q_{{\ell m }}=\left.\oint d \mathbb S^2 \operatorname{Im} \Phi_1^0(u, \theta, \phi) Y_{\ell m}(\theta, \phi)\right|_{-\infty} ^{+\infty}, \quad \quad \operatorname{Im} \Phi_1^0(u, \theta, \phi)=\lim _{r \rightarrow \infty}\left[r^2 B_r\left(u,r, \theta, \phi\right)\right] \, .
\eea
Because magnetic charges have not been observed, we do not have, in principle, a magnetic analog of a boosted charge. To the best of our knowledge,  there are no  examples of magnetic memory reported in the literature.  We discuss here one such example which, although it could be challenging to materialize physically \cite{W14}, it certainly contains pedagogical value. 
 
 To think in potential situations that exhibit magnetic memory it may be useful  to rewrite the  radial component of the magnetic field of a moving particle (\ref{magneticLW}) as
\bea
B^a(t,\vec r)\nabla_a r = \left. \frac{q / m}{4 \pi\left(1-{\vec n_s} \cdot {\vec v_s}\right)^3\left\|\vec {r}-\vec{r}_s\left(t_r\right)\right\|^2}\left\{\left({\vec n_s} \cdot {\vec a_s}\right)  L_r +\left(1-{\vec n_s} \cdot {\vec v_s}\right) \dot{ L}_r+\frac{\left(1-v_s^2\right)  L_r}{\left\|\vec{r}-\vec{r}_s\left(t_r\right)\right\|}\right\}\right|_{t_r} \, , \label{radialB}
\eea 
where $\vec a_s=\dot {\vec v}_s$ is the acceleration of the charge, and $L_r$ is the radial component of the particle's angular momentum  $\vec L=m\vec r_s\times \vec v_s$. Equivalently, these terms are related to the magnetic dipole moment of the moving charge:
\bea
\vec{m}(t)=\int_{\mathbb{R}^3} \vec{j}(t) \times \vec{r} d^3 \vec{r}=q \int_{\mathbb{R}^3} d^3 \vec{r} {\delta^{(3)}\left(\vec{r}-\vec r_s(t)\right)} \vec{v_s}(t) \times \vec{r}=q\, {\vec v_s}(t) \times {\vec r_s}(t)=-\frac{q}{m} \vec{L}(t)\, .
\eea
Equation (\ref{radialB}) shows, in particular, that for a charged particle moving with constant velocity, ${\rm Im}\Phi_1^0=0$, and all soft charges of magnetic type are zero. 

Only the first two terms in (\ref{radialB}) may lead to ${\rm Im}\Phi_1^0\neq 0$, as the third term decays as $O(r^{-3})$. Furthermore, a priori one would expect that physically reasonable situations demand $a_s(u\to \pm \infty)\to 0$. So, for sources consisting of a moving charged particle, to get magnetic memory one needs a situation in which the particle acquires a permanent rate of change for the radial angular momentum  at infinity, $\dot{ L}_r\neq 0$. 

One can imagine a situation in which this occurs. Consider polar coordinates $\{t,\rho,\phi,z\}$ and suppose there exists a non-vanishing magnetic field in $z\in(0,z_0)$ that has the  profile  $\vec B(t,\rho,\phi,z)=B_0/\rho\, \hat z$  for $\rho>\rho_0$, for some constants $B_0$ and $\rho_0>0$. Due to the inhomogeneous magnetic field, a charged particle that initially is at rest at some point in the interval $(0,z_0)$ and $\rho>\rho_0$,  and suffers a ``kick'' at some instant of time, will start spiraling outwards. It can be expected then that $\dot{ L}_r\neq 0$ at infinity. 

This problem can be solved in closed form as follows. If the kinematical variables of the charged particle are
\bea
\vec r_s & = & \rho_s \vec u_{\rho}\, ,\\
\vec v_s & = & \dot\rho_s \vec u_{\rho}+\rho_s \dot\phi_s \vec u_{\phi}\, ,\\
\vec a_s & = & (\ddot \rho_s-\rho_s \dot \phi_s ^2)\vec u_{\rho} +   (\rho_s \ddot \phi_s +2\dot \rho_s \dot \phi_s  )\vec u_{\phi}\, ,
\eea
then one can solve
\bea
\dot \rho_s(t) & = & +\sqrt{v^2-\left(\frac{q B_0}{m}+\frac{c}{\rho_s(t)} \right)^2}\, ,\\
\dot \phi_s (t) & = & \mp \frac{q B_0}{m \rho_s(t)}+\frac{c}{\rho^2_s(t)}\, ,\\
\frac{d}{dt}(\rho_s^2(t)\dot\phi_s (t)) & = & \frac{q B_0}{m} \dot \rho_s(t)\, ,
\eea
where $c$ is a constant of integration. Since $\dot \rho_s(t)>0$, $\rho_s(t)$ is monotonically increasing, therefore $\rho_s(t\to \infty)\to +\infty$ and consequently $\dot\phi_s\to 0$. The angular momentum  is simply given by $\vec {L}= \rho_s^2(t)\dot\phi_s (t)\, \vec u_z$, so
\bea
\dot L_r(t)= \frac{q B_0}{m} \dot \rho_s(t)\cos\theta\, .
\eea 
One can further check that $\vec a_s(t\to\infty)\to \vec 0$.

 Taking into account that $t_r(t,r\to\infty)\to u$, and also $\hat r\cdot \vec v_s(t) =  \dot \rho_s(t) \sin\theta \cos(\phi-\phi_s(t))+\rho_s (t)\dot\phi_s(t) \sin(\phi-\phi_s(t))$,  the limit $r\to \infty$ keeping $\{u,\theta,\phi\}$ constant gives $\left\|\vec {r}-\vec{r}_s\left(t_r\right)\right\| \sim \left\|\vec {r}-\vec{r}_s\left(u\right)\right\|\sim r$ and ${\vec n_s(t_r)} \cdot {\vec v_s(t_r)} \sim {\vec n_s(u)} \cdot {\vec v_s(u)}\sim \hat r\cdot \vec v_s(u)$. From (\ref{radialB}) we get
 \bea
 B_r(u,r,\theta,\phi)= \frac{q/m}{4\pi r^2(1- \dot \rho_s(u) \sin\theta \cos(\phi-\phi_s(u))-\rho_s (u)\dot\phi_s(u) \sin(\phi-\phi_s(u)))^2}  \frac{q B_0}{m} \dot \rho_s(u)\cos\theta +O(r^{-3})\, .
 \eea
 At future timelike infinity we have $\dot\phi_s(u\to \infty)\to 0$, $\phi_s(u\to \infty)\to \phi_0$ $\dot \rho_s(u\to \infty)\to \sqrt{v^2-\frac{q^2 B_0^2}{m^2}}$. Taking into account all this:
 \bea
 {\rm Im}\, \Phi_1^0(u\to \infty)=   \frac{q/m}{4\pi \left(1- \sqrt{v^2-\frac{q^2 B_0^2}{m^2}} \sin\theta \cos(\phi-\phi_0)\right)^2}  \frac{q B_0}{m} \sqrt{v^2-\frac{q^2 B_0^2}{m^2}}\cos\theta \, .\label{ex2}
 \eea

To finish  this section, recall from the discussion below (\ref{electricq}) that ${\rm Re} \,\Phi_1^0(u,\theta,\phi)=\Delta \alpha(u,\theta,\phi)+G(\theta,\phi)$ and ${\rm Im}\, \Phi_1^0(u,\theta,\phi)=\Delta \beta(u,\theta,\phi)$. Therefore, from (\ref{ex1}), obtained in the first example, it is possible to obtain a non-zero value of $\alpha(u,\theta,\phi)$ for $u\to \infty$, as well as $ {\rm Re}\,\Phi_1^0(\infty, \theta, \phi)- {\rm Re}\,\Phi_1^0(-\infty, \theta, \phi)\neq 0$; while from (\ref{ex2}), obtained in the second example, it is possible to obtain a non-zero value of $\beta(u,\theta,\phi)$   when $u\to +\infty$, and $ {\rm Im}\,\Phi_1^0(\infty, \theta, \phi)- {\rm Im}\,\Phi_1^0(-\infty, \theta, \phi)\neq 0$. Combining the two examples, it is not difficult  to check that (\ref{magneticq})-(\ref{electricq}) are both non-vanishing.

\section{The chiral anomaly induced by a gravitational background} \label{sec5}

As remarked in the Introduction, if instead of an electromagnetic field we consider an external gravitational background, described by a curved spacetime ($M$, $g_{ab}$), Dirac fields (as well as the  electromagnetic field itself \cite{AdRNS16, AdRNS17,AdRNS18a,AdRNS18b}) experience a gravitationally-induced chiral anomaly (see Eqn.~\eqref{totalanomaly}).  Similarly to what we did for electromagnetic backgrounds, we  explore here global properties of this chiral anomaly by studying the change of the chiral charge $Q_A$.

As usual in General Relativity, we restrict to globally hyperbolic spacetimes to ensure the well-posedness of the Cauchy problem. This allows us to foliate the manifold in the form $M\simeq \mathbb R \times \Sigma$. We will further assume that the spatial slices are $\Sigma \simeq \mathbb R^3$. Performing a similar analysis as in (\ref{integral}), the permanent change in the chiral charge predicted by the chiral anomaly is dictated now by the  Chern-Pontryagin integral 
\bea   \left< Q_A\right>_{\mathcal J^{+}}- \left< Q_A \right>_{\mathcal J^{-}}=\frac{\hbar}{192\pi^2}  \int_{\mathbb R^4}d^4x \sqrt{-g}  R_{abcd}\, ^{\star}R^{abcd}\, ,
\eea
where $\{x^a\}$ is a global coordinate system for $M\simeq \mathbb R^4$. The RHS of this equation was investigated in  \cite{dR21, dRetal2020}. Although it may appear intractable from an analytical viewpoint, it is actually possible to rewrite it in a form that allows us to extract information of  physical value without having to resort to numerical techniques. More precisely, assuming no incoming gravitational waves from past null infinity $\mathcal J^-$, it is possible to rewrite it as an integral over future null infinity only:
\bea
 \left< Q_A\right>_{\mathcal J^{+}}- \left< Q_A \right>_{\mathcal J^{-}} =-\frac{\hbar}{96 \pi^2} \int_{-\infty}^{\infty} d u \int d \mathbb S^2 {\rm Im}(N \bar\Psi_4 )\, . \label{gravity}
\eea

In this expression, $\Psi_4(u,\theta,\phi)=-\lim_{r\to \infty} r C_{abcd}\bar m^a n^b \bar m^c n^d$ is a complex scalar constructed from the Weyl tensor $C_{abcd}$, which carries the two radiative degrees of freedom of gravitational waves; it is the gravitational analog of the  complex scalar $\Phi_2^0(u,\theta,\phi)$ in electrodynamics (compare with equations (\ref{phi2}) and (\ref{phi20})). On the other hand,  $N(u,\theta,\phi)=N_{ab}(u,\theta,\phi)m^a m^b$ is the  relevant component of the Bondi News tensor $N_{ab}$ \cite{G77}, which measures the time evolution of the asymptotic shear  of outgoing null geodesics at $\mathcal J^+$. It is a symmetric, transverse ($N_{ab}n^b=0$) and traceless tensor on $\mathcal J^+$ that, just like  $\Psi_4$,   captures the two gravitational degrees of freedom at future null infinity. The two quantities are related by  $\Psi_4=-\frac{1}{2}\dot N$, so $N$ can be thought of as the gravitational analog of the electromagnetic potential $A_2^0$ (compare with equation (\ref{phi22}) with the gauge choice $A_1^0=0$).  The total amount of energy  carried away by the gravitational waves across $\mathcal J^+$ is proportional to $\int_{-\infty}^{\infty}d\mathbb S^2 du |N(u,\theta,\phi)|^2$. Because of this, the Bondi News indicates unambiguously if a system is radiating gravitational waves. If $N=0$ then the sources do not emit radiation, while $N\neq 0$ indicates the presence of radiation. Finiteness of this energy flux requires $N(\cdot, \theta,\phi)\in L^2(\mathbb R, \mathbb C)$ for all $(\theta,\phi)\in \mathbb S^2$, and in particular $N\to 0$ as $u\to \pm \infty$. These properties carry over to $\Psi_4$.

In view of the results found in Sec. \ref{sec4}, it is natural to ask if gravitational soft charges, or gravitational memory, may also contribute to the fermion chiral anomaly (\ref{gravity}).
The gravitational memory effect \cite{ZP74, BG85, BT87, C91, T92, F92, BG14} consists in the permanent relative displacement that a set of free test  masses may experience after the passage of a gravitational wave burst. The deformation of a congruence of free observers or curves is controlled by the shear. If $\sigma(u,\theta,\phi)$ denotes the asymptotic shear of outgoing  null geodesics at future null infinity, a flux of gravitational radiation will make $\sigma(u,\theta,\phi)$ evolve with time $u$, while it remains constant otherwise. As commented above, this effect is  captured precisely in the Bondi News, which is related to the shear  via the equation $N=2 \dot \sigma$. Because $N\to 0$ as $u\to \pm \infty$, $\sigma(u,\theta,\phi)$ reaches constant values at early and late times. However, $\sigma(-\infty,\theta,\phi)\neq \sigma(\infty,\theta,\phi)$ in general, and there can remain a permanent distortion in the shear. The amount of gravitational memory encoded in free test masses is quantified then by the overall change in the asymptotic shear $\sigma(u,\theta,\phi)$ of outgoing null geodesics between early and late times:
\bea
q_\alpha= \frac{1}{8\pi} \int d\mathbb S^2 \, \bar\eth^2 \alpha(\theta,\phi) (\sigma(\infty,\theta,\phi)-\sigma(-\infty,\theta,\phi))\, ,
\eea
where $\alpha$ is an arbitrary real-valued function on the sphere. These quantities are called gravitational infrared charges  \cite{A14} (compare this definition with  the electromagnetic analog (\ref{softcharges})). Following the analogy with the electromagnetic case, it can also be proven that these charges can be identified with the Hamiltonian generating BMS supertranslations in the radiative phase space of General Relativity \cite{A81a, A81b}. From the point of view of the bulk, supertranslations are diffeomorphisms (the gauge transformations in General Relativity) that do not vanish at infinity, as a result of which  they are  called ``large''.

Notice that the physical manifestations of the gravitational and electromagnetic memory effects are qualitatively different. An electromagnetic field does not generate a permanent, relative displacement of electrically charged particles; instead, it generates a permanent, relative velocity between the charges.

Using the relation between the shear and the Bondi news,  we can  formulate the gravitational infrared charges in terms of the radiative degrees of freedom,
\bea
q_\alpha=\frac{1}{16\pi}\int_{-\infty}^{+\infty} du d\mathbb S^2 N(u,\theta,\phi) \bar\eth^2 \alpha(\theta,\phi) \, .
\eea
Expanding in a basis of spin-weighted spherical harmonics, this expression reduces to 
\bea
q_\alpha=\frac{1}{16\pi}\int_{-\infty}^{+\infty} du \sum_{\ell m} \alpha_{\ell m} N_{\ell m}(u) \equiv \frac{1}{16\pi} \sum_{\ell m}\alpha_{\ell m} q_{\ell m}\, ,
\eea
for  real-valued coefficients $\alpha_{\ell m}$. In the second equality, we have defined the parameters $q_{\ell m}=\int_{-\infty}^{+\infty} du N_{\ell m}(u)$.
Now, because $N(\cdot, \theta,\phi)\in L^2(\mathbb R, \mathbb C)$, each of its harmonic modes admits a Fourier transform on $\mathcal J^+$
\bea
\tilde N_{\ell m}(\omega)=\int_{-\infty}^{+\infty} d u\, N_{\ell m}(u) \, e^{-i \omega u} \, . 
\eea
Therefore, just like in the electromagnetic case,  we conclude that the infrared charges are determined by the zero-frequency mode of the gravitational radiation {\it as described by the Bondi News $N$}.  Namely,
\bea
q_\alpha=\frac{1}{16\pi} \sum_{\ell m}\alpha_{\ell m} \tilde N_{\ell m}(0)\, .
\eea

Notice, however, that, in sharp contrast with the electromagnetic case (\ref{softcharges}), the infrared charges are determined by the zero modes of the ``potential'' $N(u,\theta,\phi)$  and not by the zero modes of the ``field'' $\Psi_4(u,\theta,\phi)$. While this may look an irrelevant comment, it is an important point in our analysis. The calculation of the RHS of (\ref{gravity}) is formally equal to the electromagnetic case (\ref{electromagnetism}) if we identify $A_2^0$ with $N$, and $\Phi_2^0$ with $\Psi_4$. In the previous section we found that the electromagnetic infrared charges contribute to the chiral anomaly through the zero-modes of the electromagnetic field $\Phi_2^0(u,\theta,\phi)$. Similarly, in the gravitational case, (\ref{gravity}) only receives  contributions from the zero modes of $\Psi_4^0(u,\theta,\phi)$, while the zero mode of $N(u,\theta,\phi)$ never appears. However,   $\Psi_4^0(u,\theta,\phi)$ has no zero mode:
\bea
\tilde\Psi_4(0,\theta,\phi) = \int_{-\infty}^{+\infty} du \Psi_4(u,\theta,\phi) = -\frac{1}{2} \left[ N(+\infty,\theta,\phi) -  N(-\infty,\theta,\phi)\right] = 0\, ,
\eea
where in the second equality we made use of  $\Psi_4=-\frac{1}{2}\dot N$,  and the last equality follows from $N(\pm\infty,\theta,\phi)=0$. This is in sharp contrast with electrodynamics, where $\Phi_2^0(u,\theta,\phi)$ ---the electromagnetic analog of $\Psi_4$---  does have a zero mode. As a consequence, only the radiative part of the gravitational field contributes to the chiral fermion anomaly in (\ref{gravity}). There is no gravitational memory contributing to the change  of the chiral charge $ \left< Q_A\right>$, and the total change from $\mathcal J^-$ to $\mathcal J^+$ is determined by the helicity carried away by gravitational waves generated in the bulk \cite{dRetal2020, dR21}.

\section{Conclusions} \label{sec6}

Chiral fermion anomalies have been extensively studied in the literature for several decades and from multiple viewpoints. Despite that, this topic is sufficiently rich to allow for yet another intriguing insight. We have found one such new aspect by studying global aspects of the chiral anomaly, related to the failure of the chiral charge $Q_A$ of a massless Dirac field to be conserved. This charge is strictly conserved classically, as well as in quantum field theory for free Dirac fields. However, the presence of background fields, either electromagnetic or gravitational, may induce a local non-conservation of the chiral current $j_A^a$ by quantum fluctuations, which can potentially produce a time evolution in the vacuum expectation value $\langle Q_A\rangle$. 

The identification of  external fields that can or cannot trigger a change of $\langle Q_A\rangle$ is a non-trivial problem. For non-abelian gauge fields, a traditional approach is to look for instanton solutions in an euclidean spacetime, which display a complex topological/global structure. To address this question,  we have evaluated instead the change in $\langle Q_A\rangle$ between past and future null infinity using familiar, global techniques within the framework of asymptotically flat spacetimes.
For an external electromagnetic field, our results are neatly summarized in equation (\ref{main}). This equation tells us that $\langle Q_A\rangle$ can change between past and future null infinity if (i) electromagnetic sources in the bulk emit circularly polarized electromagnetic waves  (i.e. radiation with net helicity) and/or (ii) if electromagnetic sources in the bulk produce  transitions between certain infrared sectors of Maxwell theory. The relevant transitions are determined by a concrete pair of  infrared charges of electric and magnetic type, respectively, written in equation \eqref{magneticq} and \eqref{electricq}. To gain physical intuition, we have devised an academic example where the required soft charges are different from zero.

Physically, non-zero infrared charges are known to produce memory effects on physical systems.  This is how the transitions between the infrared quantum vacua can leave observable imprints. To the best of our knowledge, the   only electromagnetic memory effects  known to date involve  classical systems. Here, we have shown that quantum states of a field theory can also keep memory of the past influence of  electromagnetic radiation, by storing a certain amount of helicity. 

The connection of electromagnetic memory and the change of $\langle Q_A\rangle$ has also been worked out in 1+1 dimensions, which is cleaner because there are no electromagnetic waves. This example also allowed us to interpret this new memory effect  in terms of ``kicks'' of virtual charges and excitation of particle pairs out of the quantum vacuum.

Overall, the results in this paper, together with our previous analysis  \cite{dR21,dRetal2020}, open up  an unforeseen connection between chiral  anomalies, the radiative content of the electromagnetic field, infrared charges and the memory effect. 

Although our approach is qualitatively different, the contribution from soft charges to the chiral anomaly bears some similarity with the rationale underlying instantons in Euclidean gauge-field theories. According to the usual interpretation \cite{J77, S94}, instantons mediate quantum-mechanical transitions between inequivalent vacuum states of the Hilbert space of the background (non-abelian) gauge field.  These transitions, which are labeled by the instanton charge, are able to induce the chiral anomaly \cite{TH76}. 
On the other hand, the quantization of the electromagnetic field at future null infinity leads naturally to a Hilbert space that can be divided in different, disjoint infrared sectors \cite{AN81, A87}, which represent inequivalent notions of quantum vacua. The infrared charges label  transitions between the different infrared sectors, and therefore play the same role of the instanton charge. We have shown in this article that these transitions contribute to the chiral anomaly in a specific manner.  

To finish, we have also checked that, quite interestingly, gravitational infrared charges do not contribute to the fermion chiral anomaly.

\section{Acknowledgements}

 I.A.\  thanks Miguel Campiglia for numerous illuminating discussions on the content of this article. ADR is grateful to A. Ashtekar for a deep training of many of the techniques employed in this article.
ADR is supported through a M. Zambrano grant (ZA21-048) with reference UP2021-044 from the Spanish Ministerio de Universidades, funded within the European Union-Next Generation EU. This work is also supported by the Spanish Grant PID2020-116567GB-C21 funded by MCIN/AEI/10.13039/501100011033, the project PROMETEO/2020/079 (Generalitat Valenciana), as well as by the NSF grant PHY-1806356, and the Eberly Chair funds of Penn State.
I.A.\ is supported by the NSF grant PHY-2110273, by the RCS program of  Louisiana Boards of Regents through the grant LEQSF(2023-25)-RD-A-04, by the Hearne Institute for Theoretical Physics, and in part by Perimeter Institute for Theoretical Physics. Research at Perimeter Institute is supported by the Government of Canada through the Department of Innovation, Science, and Economic Development, and by the Province of Ontario through the Ministry of Colleges and Universities.
\appendix

\section{Chiral fermion anomaly and  spontaneous particle creation in 1+1 dimensions} \label{particlecreation2d}

In Section \ref{sec2} we argued that the permanent change in the vacuum expectation value  $\langle Q_A\rangle$ produced by the chiral anomaly  (\ref{totalanomaly}) can be understood as a net creation of helicity resulting from virtual particles that are ``kicked'' out of the vacuum state. In 1+1 dimensions it is  relatively easy  to see the connection between the chiral fermion anomaly and particle pair creation. In this appendix we show this connection and deduce the anomaly from the analysis of Bogoliubov transformations between canonical in and out vacua. We perform the analysis for a general, non-uniform background electric field $E(t,x)$. 

When the external electric field is uniform, $E=E(t)$, a heuristic argument involving the Dirac ``sea''  has been  given in several occasions in the literature (see \cite{J85, NN83, AGP83, AGM12}). However, to the best of our knowledge, an explicit and/or rigorous, complete calculation is still lacking. In particular, this heuristic picture always misses a contribution from vacuum polarization, which we provide here. 

Let us consider a massless, quantum Dirac field $\Psi(t,x)$ living in a two-dimensional flat spacetime, $(\mathbb R\times \mathbb S^1,\eta_{ab})$, and coupled to an external electric field $E(t,x)$. The spatial sections will have length equal to $L$. The electric field  departs from an initial, ``vacuum''  configuration at early times, $E(t_{\rm in},x)=0$, it is switched on for a finite amount of time, and eventually returns  to another  ``vacuum'' state, $E(t_{\rm out},x)=0$. At early and late times, in which the field strength vanishes, one can introduce canonical ``in'' and ``out'' vacuum states for the fermion field. As remarked in Sec. \ref{sec2}, the two electric vacua are equivalent in the classical theory, but they may differ in the quantum theory if the potential $A_a(t,x)$ changes non-trivially, as a result of which the ``out'' state of the fermion field will potentially differ from the ``in'' state. 

As in Section \ref{sec2}, we  work with the temporal gauge fixing:  $A_t(t,x)=0$. This can always be obtained by a suitable gauge transformation. There is still a residual gauge freedom, which consists in $A_a(t,x) \to A_a(t,x) + \nabla_a \alpha$, for $\alpha=\alpha(x)$. This residual freedom can be fixed by demanding $A_a(t\to -\infty,x)\to 0$, which we will adopt here onwards.
At early and late times, where $E(t,x)=0$, we have $\partial_t A_x=\partial_x A_t$, and the electromagnetic connection 1-form takes the form  $A(t\to -\infty,x)=0$, and  $A(t\to \infty,x)=A(x)dx$.

Massless Dirac fields $\Psi(t,x)$ split into two decoupled, left- and right-handed Weyl spinors, that will be denoted by $u_+$, $u_-$, respectively. In 1+1 dimensions these spinors are represented by ordinary functions on the spacetime. For $t<t_{\rm in}$, the two Weyl equations read:
\bea
\begin{array}{l}
\left(i \partial_t-i \partial_x\right) u_{+}=0\, , \\
\left(i \partial_t+i \partial_x\right) u_{-}=0\, ,
\end{array}
\eea
while for $t>t_{\rm out}$ we have
 \bea
\left(i \partial_t-i \partial_x-qA(x)\right) u_{+}&=&0\, , \label{Eplus}\\
\left(i \partial_t+i \partial_x+qA(x)\right) u_{-}&=&0\, ,\label{Eminus}
\eea
where the evolution of the electric field has left a residual gauge potential in the field equations.
To study the effects of the external, dynamical electric  field on the  fermion modes, it is useful to analyze the Bogoliubov transformations between the  in an out vacuum states. To specify these states we have to provide first a  basis of positive-frequency solutions for the in and out  Hilbert spaces. 
 
Let us focus on equations (\ref{Eplus})-(\ref{Eminus}) (the solutions at early times can be simply recovered by taking $A=0$).
These equations admit separable solutions of the form $u_{\pm}(t,x)=f_{\pm}(t)g_{\pm}(x)$, which produce
\bea
\frac{i \partial_t f_{\pm}(t)}{f_{\pm}(t)}=\pm\frac{\left( \partial_x+qA(x)\right) g_{\pm}(x)}{g_{\pm}(x)} \equiv \omega_{\pm}={\rm const}\, ,
\eea 
for some separation constants $\omega_{\pm}$.
The LHS of this equation can be solved to give $f_{\pm}(t)=e^{-i\omega_{\pm} t}$, and are of positive frequency  with respect to the operator $H=i\partial_t$ if $\omega_{\pm}>0$. To solve the spatial dependence of the field modes, note that the potential must be pure gauge (because $E=0$), so let us write $qA=\partial_x\phi(x)$ for some function $\phi(x)$. Then $\left({i} \partial_x+qA(x)\right) g_t(x)=e^{+i \phi(x)} i \partial_x\left(e^{-i \phi(x)}g_{\pm}\right)$ and the equation above yields $g_{\pm}(x)=e^{\mp i \omega_{\pm} x}e^{i\phi(x)}$. In conclusion, a basis of  positive-frequency solutions to the Weyl equations above consist  of $\{u_{\pm,\omega}(t,x)\}_{\omega>0}$,  where
\bea
u_{\pm,\omega}(t,x)=e^{-i\omega_{\pm} (t\pm x)}e^{ i\phi(x)}\, .
\eea
These functions represent left- and right- moving  modes, respectively, and they define the two chiral sectors of the theory. In 1+1 dimensions handedness is nothing but the direction of propagation in the  spatial dimension.

At early times the connection is identically zero, $A=0$, so $\phi(x)=0$ for the in modes: $u^{{\rm in}}_{\pm,\omega}(t,x)=e^{-i\omega_{\pm} (t\pm x)}$. However, the dynamics of the electric field could produce $\phi(x)\neq 0$ at late times. This is the electric memory mentioned in the main text. The out modes can differ  by a position-dependent, global phase with respect to the in modes: $u^{{\rm out}}_{\pm,\omega}(t,x)=e^{-i\omega_{\pm} (t\pm x)}e^{ i\phi(x)}$. 

The compactness of the spatial spacetime dimension imposes severe constraints on  the field modes.
 On $\mathbb S^1$ there are two inequivalent spin structures: the trivial one (that corresponds to imposing periodic boundary conditions on the field modes), and the non-trivial one (that corresponds to imposing anti-periodic boundary conditions) \cite{parker-toms}. That is, $u_{\pm,\omega}(t,x+L)=e^{i2\pi \delta} u_{\pm,\omega}(t,x)$, where $\delta=0$ or $1/2$ for periodic and anti-periodic boundary conditions, respectively. We shall assume here periodic boundary conditions, for simplicity. This implies
$
e^{\mp i \omega_{\pm} L} e^{i \phi(x+L)}=e^{i \phi(x)}\, ,
$
which produces $\mp \omega_{\pm,n} L+\phi(x+L)-\phi(x)=-n 2 \pi,   n\in \mathbb{Z}$, or
\bea
\omega_{\pm,n}=\pm n \frac{2 \pi}{L} \pm \frac{(\phi(x+L)-\phi(x))}{L}\, .
\eea
In other words, the allowed  frequencies of the field modes take only discrete values. To convert the RHS in terms of the electromagnetic potential, recall that $qA_x=\partial_x \phi(x)$, so $\phi(x)=q\int^x A(x')dx'$, and $\phi(x+L)-\phi(x) = q\int_x^{x+L}A(x')dx'$. This integral is well-defined, so we can make a change of variable $x' \to x' - x$ to rewrite it as
$
\phi(x+L)-\phi(x) = q\int_0^L A(x')dx' =: 2\pi q\, CS[A]
$, where $CS[A]$ is the Chern-Simons \cite{Nakahara}. Therefore,
\bea
\omega_{\pm,n}=\pm\frac{2\pi}{L}(n+q\, CS[A])\, .
\eea
From this result we can infer,  
\bea
\omega_{\pm,n}>0\quad {\rm implies} \quad \left\{\begin{array}{ll}n>-q\,C S[A] & \text { for positive chirality }\, , \\ n<-q\, C S[A] & \text { for negative chirality }\, ,\end{array}\right. \\
\omega_{\pm,n}<0 \quad {\rm implies} \quad \left\{\begin{array}{ll}n<-q\,C S[A] & \text { for positive chirality } \, ,\\ n>-q\,C S[A] & \text { for negative chirality }\, .\end{array}\right. 
\eea 

The field equations (\ref{Eplus})-(\ref{Eminus}) are linear, and therefore the space of solutions of each chiral sector has the structure of a vector space. As usual,  we endow these vector spaces with the Dirac inner product $(u_1,u_2)=\int_0^L dx \bar u_1(t,x) u_2(t,x)$, which is preserved in time by the Weyl equations and by the periodic boundary conditions.  Then, an orthonormal basis $f^{{\rm in}}_{\pm, n}(t,x)$ of periodic, positive-frequency modes for the in Hilbert space $L^2((0,L))\oplus L^2((0,L))$ is given by $\{\frac{e^{-i\frac{2\pi}{L}n (t+ x)}}{\sqrt{L}}\}_{n\in \mathbb Z^+}$ for left-moving spinors and $\{\frac{e^{i\frac{2\pi}{L}n (t- x)}}{\sqrt{L}}\}_{n\in \mathbb Z^-}$ for right-moving spinors. If we define the helicity as $h:=\hbar\frac{\omega_{\pm,n}}{\frac{2\pi}{L}n}=\pm \hbar$, we see that for left-moving spinors, $u_+$, positive-frequency modes have positive helicity, while for right-moving spinors $u_-$ positive-frequency modes have negative helicity ($\frac{2\pi}{L}n$ can be thought of as the wave-number of the mode). On the other hand, an orthonormal basis $f^{{\rm out}}_{\pm, n}(t,x)$ of periodic, positive-frequency modes for the out Hilbert space $L^2((0,L))\oplus L^2((0,L))$ is given, at late times, by 
\bea
\left\{\frac{ e^{-i\frac{2\pi}{L}n (t+ x)} }{\sqrt{L}} e^{i \frac{2\pi}{L} q\,CS[A] (- t-x) +iq \int^x A(x')dx'}\right \}_{n+q\,CS[A]\in \mathbb Z^+}
\eea
 for left-moving spinors, and 
 \bea
 \left\{\frac{ e^{i\frac{2\pi}{L}n (t- x)} }{\sqrt{L}} e^{i \frac{2\pi}{L} q\,CS[A] ( t-x) +iq\int^x A(x')dx'}\right\}_{n+q\,CS[A]\in \mathbb Z^-}
 \eea
   for right-moving spinors. 

In the basis of in modes  the quantum fields can be expanded as
\bea
u_{+}(t,x)=\sum_{n>0}^{+\infty} a^{{\rm in}}_{+, n} f^{{\rm in}}_{+, n}(t,x)+ \sum_{n<0}^{-\infty} b_{-,n}^{{\rm in}\dagger} {f^{{\rm in}}_{+, n}(t,x)} \, , \label{inplus}\\
u_{-}(t,x)=\sum_{n<0}^{-\infty} a^{{\rm in}}_{-, n} {f^{{\rm in}}_{-, n}}(t,x)+ \sum_{n>0}^{+\infty} b_{+,n}^{{\rm in}\dagger} {f^{{\rm in}}_{-, n}}(t,x) \, , \label{inminus}
\eea
where $ f^{{\rm in}}_{\pm, n}(t,x) \to \frac{e^{\mp i\frac{2\pi}{L}n (t\pm x)} }{\sqrt{L}}$ at early times. In this expression $a^{{\rm in}}_{\pm, n}$ annihilate fermions (negative charges) with positive (+) and negative (-) helicity, while $b^{{\rm in}\dagger}_{\mp, n}$ create anti-fermions (positive charges) with negative (-) and positive (+) helicity. The $\pm$ signs on $u_{\pm}$ and $f^{{\rm in}}_{\pm, n}$ simply denote the sign of chirality (i.e. the direction of propagation). On the other hand, in the basis of out modes we have instead
\bea
u_{+}(t,x)=\sum_{n>-q\,CS[A]}^{+\infty} a^{{\rm out}}_{+, n} f^{{\rm out}}_{+, n}(t,x)+ \sum_{n<-q\,CS[A]}^{-\infty} b_{-,n}^{{\rm out}\dagger} {f^{{\rm out}}_{+, n}(t,x)}\, , \label{outplus}\\
u_{-}(t,x)=\sum_{n<-q\,CS[A]}^{-\infty} a^{{\rm out}}_{-, n} {f^{{\rm out}}_{-, n}}(t,x)+ \sum_{n>-q\,CS[A]}^{+\infty} b_{+,n}^{{\rm out}\dagger} {f^{{\rm out}}_{-, n}}(t,x) \, ,\label{outminus}
\eea
where $ f^{{\rm out}}_{\pm, n}(t,x) \to \frac{ e^{\mp i\frac{2\pi}{L}n (t\pm x)} }{\sqrt{L}} e^{i \frac{2\pi}{L} q\,CS[A] (\mp t-x) +iq\int^x A(x')dx'}$ at late times. The Bogoliubov coefficients that relate the in- and out-representations are now  easy to obtain. Since  $f^{{\rm out}}_{\pm, n}(t,x)$ form a complete basis of  the Hilbert space of solutions, $ f^{{\rm in}}_{\pm, n}(t,x)$ can be expanded as
\bea
n>0 \, \, ({\rm positive\,\,  energy}): \quad && f^{{\rm in}}_{+, n}(t,x)  =  \sum_{n'>-q\,CS[A]}^{+\infty} \alpha^+_{n n'} f^{{\rm out}}_{+, n'}(t,x)+ \sum_{n'<-q\,CS[A]}^{-\infty} \beta^+_{nn'} {f^{{\rm out}}_{+, n'}(t,x)} \, ,\label{positivein+}\\
n<0  \, \, ({\rm negative\,\,  energy}): \quad && f^{{\rm in}}_{+, n}(t,x)  =  \sum_{n'<-q\,CS[A]}^{-\infty} \tilde\alpha^+_{n n'} f^{{\rm out}}_{+, n'}(t,x)+ \sum_{n'>-q\,CS[A]}^{+\infty} \tilde\beta^+_{nn'} {f^{{\rm out}}_{+, n'}(t,x)} \, , \label{negativein+}\\
 n<0  \, \, ({\rm positive\,\,  energy}):  \quad&& f^{{\rm in}}_{-, n}(t,x)  =  \sum_{n'<-q\,CS[A]}^{-\infty}  \alpha^-_{n n'} {f^{{\rm out}}_{-, n'}}(t,x)+ \sum_{n'>-q\,CS[A]}^{+\infty} \beta^-_{nn'} {f^{{\rm out}}_{-, n'}}(t,x) \, ,\label{positivein-}\\
 n>0  \, \, ({\rm negative\,\,  energy}):  \quad&& f^{{\rm in}}_{-, n}(t,x)  =  \sum_{n'>-q\,CS[A]}^{+\infty}  \tilde\alpha^-_{n n'} {f^{{\rm out}}_{-, n'}}(t,x)+ \sum_{n'<-q\,CS[A]}^{-\infty} \tilde\beta^-_{nn'} {f^{{\rm out}}_{-, n'}}(t,x) \, . \label{negativein-}
\eea
Using the normalization conditions  $(f_{\pm,n}, f_{\pm,n'})=\delta_{n,n'}$ one can get several useful identities. For instance:
\bea
\delta_{n n'}=(f^{{\rm in}}_{+, n}, f^{{\rm in}}_{+, n'}) =  \sum_{n''>-q\,CS[A]}^{+\infty}\bar  \alpha^+_{n n''}  \alpha^+_{n' n''}+ \sum_{n''<-q\,CS[A]}^{-\infty}\bar  \beta^+_{n n''}  \beta^+_{n' n''} \, , \label{id1}\\
\delta_{n n'}=(f^{{\rm in}}_{-, n}, f^{{\rm in}}_{-, n'}) =  \sum_{n''<-q\,CS[A]}^{-\infty}\bar  \alpha^-_{n n''}  \alpha^-_{n' n''}+ \sum_{n''>-q\,CS[A]}^{+\infty}\bar  \beta^-_{n n''}  \beta^-_{n' n''}  \, . \label{id2}
\eea
Setting $n=n'$, the convergence of the integrals imply that $\lim_{n''\to+\infty}|\beta^{\pm}_{nn''}|=0$.
In what follows we will also need the inverse Bogoliubov transformations. Since $f^{{\rm in}}_{\pm, n}(t,x)$ also form a complete basis of  the Hilbert space of solutions, the elements $ f^{{\rm out}}_{\pm, n}(t,x)$ can equivalently be expanded as
\bea
n>-q\,CS[A] \, \, ({\rm positive\,\,  energy}): \quad && f^{{\rm out}}_{+, n}(t,x)  =  \sum_{n'>0}^{+\infty} \gamma^+_{n n'} f^{{\rm in}}_{+, n'}(t,x)+ \sum_{n'<0}^{-\infty} \delta^+_{nn'} {f^{{\rm in}}_{+, n'}(t,x)} \, , \\
n<-q\,CS[A] \, \, ({\rm positive\,\,  energy}): \quad && f^{{\rm out}}_{-, n}(t,x)  =  \sum_{n'<0}^{-\infty} \gamma^-_{n n'} f^{{\rm in}}_{-, n'}(t,x)+ \sum_{n'>0}^{+\infty} \delta^-_{nn'} {f^{{\rm in}}_{-, n'}(t,x)} \, .
\eea
For $n''>0$ we have $\gamma^+_{n n''}=(f^{{\rm in}}_{+, n''},f^{{\rm out}}_{+, n})=\overline{ \alpha_{n''n}^+}$ and for $n''<0$ we have $\delta^+_{n n''}=(f^{{\rm in}}_{+, n''},f^{{\rm out}}_{+, n})=\overline{ {\tilde\beta}_{n''n}^+}$. Similarly, for $n''<0$ we obtain $\gamma^-_{n n''}=(f^{{\rm in}}_{-, n''},f^{{\rm out}}_{-, n})=\overline{ \alpha_{n''n}^-}$, while for $n''>0$ we get  $\delta^-_{n n''}=(f^{{\rm in}}_{-, n''},f^{{\rm out}}_{-, n})=\overline{ {\tilde\beta}_{n''n}^-}$. Using   $(f_{\pm,n}, f_{\pm,n'})=\delta_{n,n'}$ we further obtain the following identities
\bea
\delta_{n n'}=(f^{{\rm out}}_{+, n}, f^{{\rm out}}_{+, n'}) =  \sum_{n''>0}^{+\infty}\overline{  \alpha^+_{n'' n}}  \alpha^+_{n'' n'}+ \sum_{n''<0}^{-\infty}\overline{\tilde  \beta^+_{n'' n}}\tilde\beta^+_{n'' n'} \, , \label{id3} \\ 
\delta_{n n'}=(f^{{\rm out}}_{-, n}, f^{{\rm out}}_{-, n'}) =  \sum_{n''<0}^{-\infty}\overline{  \alpha^-_{n'' n}}  \alpha^-_{n'' n'}+ \sum_{n''>0}^{+\infty}\overline{\tilde  \beta^-_{n'' n}}  \tilde\beta^-_{n'' n'} \, .  \label{id4}
\eea
Setting $n=n'$, the convergence of the integrals imply that $\lim_{|n''|\to+\infty}|\tilde\beta^{\pm}_{n''n}|=0$.

Plugging (\ref{positivein+}) and (\ref{negativein+})  in (\ref{inplus}), and rewriting as in (\ref{outplus}), we can read off the Bogoliubov transformation of the creation and annihilation operators:
\bea
a^{{\rm out}}_{+, n'} & = & \sum_{n>0}^{+\infty} \alpha^{+}_{n n'} a^{{\rm in}}_{+, n}+ \sum_{n<0}^{-\infty}\tilde \beta^{+}_{n n'} b^{{\rm in}\dagger}_{-, n}\, , \label{op1}\\
b^{{\rm out}\dagger}_{-, n'} & = & \sum_{n<0}^{-\infty} \tilde\alpha^{+}_{n n'} b^{{\rm in}\dagger}_{-, n}+ \sum_{n>0}^{\infty} \beta^{+}_{n n'} a^{{\rm in}}_{+, n} \, . \label{op2}
\eea 
Similarly, plugging (\ref{positivein-}) and (\ref{negativein-})  in (\ref{inminus}), and rewriting as in (\ref{outminus}), we can read off the Bogoliubov transformation for the remaining creation and annihilation operators:
\bea
a^{{\rm out}}_{-, n'} & = & \sum_{n<0}^{-\infty} \alpha^{-}_{n n'} a^{{\rm in}}_{-, n}+ \sum_{n>0}^{+\infty}\tilde \beta^{-}_{n n'} b^{{\rm in}\dagger}_{+, n} \, , \label{op3}\\
b^{{\rm out}\dagger}_{+, n'} & = & \sum_{n>0}^{+\infty} \tilde\alpha^-_{n n'} b^{{\rm in}\dagger}_{+, n}+ \sum_{n<0}^{-\infty} \beta^{-}_{n n'} a^{{\rm in}}_{-, n}\, .  \label{op4}
\eea

We have now all the required ingredients  to see the connection between the chiral anomaly and particle pair production. Using the normalization equations $(f_{\pm,n}, f_{\pm,n'})=\delta_{n,n'}$,  the chiral charge can be  formally evaluated as:
\bea
Q_5(t) & = &\int_0^Ldx \bar\Psi(t,x)\gamma^0\gamma_5\Psi(t,x) = \int_0^Ldx \Psi^{\dagger}(t,x)\gamma_5\Psi(t,x)=\int_0^Ldx (\overline{u}_+ u_+ - \overline{u}_- u_- )= (u_+,u_+)-(u_-,u_-) \nonumber \\
&= & \sum_{n>-q\,CS[A]}  a_{+,n}^{\rm out\dagger} a_{+,n}^{\rm out}+\sum_{n<-q\,CS[A]}b_{-,n}^{\rm out} b_{-,n}^{\rm out \dagger}-\sum_{n<-q\,CS[A]}a_{-,n}^{\rm out\dagger} a_{-,n}^{\rm out}-\sum_{n>-q\,CS[A]} b_{+,n}^{\rm out} b_{+,n}^{\rm out \dagger}\, . \nonumber
\eea
However,  this operator is not well-defined in our Fock space, its expectation values produce  divergent sums. This is because $Q_5(t)$ is quadratic in the quantum fields, and, consequently, the evaluation of expectation values requires renormalization. The quantity of interest is $\langle {\rm in}| Q_5(t) |{\rm in}\rangle_{\rm{ren}}$, whose time evolution tells us whether there exists an anomaly or not. To obtain this result one can apply renormalization directly. However, there is an alternative, indirect procedure which, as we shall see,  provides useful insights on the physical interpretation of $\langle {\rm in}| Q_5(t) |{\rm in}\rangle_{\rm{ren}}$.
Let us  introduce the following fiducial (``normal-ordered'') operator:
\bea
:Q_5:(t_{\rm out}) &:=&  \int_0^Ldx \lim_{y\to x}\left[ \bar\Psi(t_{\rm out},x)\gamma^0\gamma_5\Psi(t_{\rm out},y) - \mathbb I  \langle {\rm out}| \bar\Psi(t_{\rm out},x)\gamma^0\gamma_5\Psi(t_{\rm out},y) |{\rm out}\rangle      \right]  \label{fiducial}\\
& = &\sum_{n>-q\,CS[A]}  a_{+,n}^{\rm out\dagger} a_{+,n}^{\rm out}-\sum_{n<-q\,CS[A]}b_{-,n}^{\rm out \dagger} b_{-,n}^{\rm out }-\sum_{n<-q\,CS[A]}a_{-,n}^{\rm out\dagger} a_{-,n}^{\rm out}+\sum_{n>-q\,CS[A]} b_{+,n}^{\rm out \dagger} b_{+,n}^{\rm out } \, . \nonumber
\eea
This operator, which  is given in terms of particle number operators of the out vacuum state, is now well-defined on the Fock space. In particular, the expectation value $\langle {\rm in}| :Q_5:(t_{\rm out}) |{\rm in}\rangle$ exists.  However, keep in mind that this is just an auxiliary operator that we introduced for convenience.  What truly determines the quantum anomaly is the time evolution of the charge $Q_5$, {\it not} of the fiducial operator $:Q_5:$. 
Using the definition above we can obtain the relation between the two:
\bea
\langle {\rm in}| :Q_5:(t_{\rm out}) |{\rm in}\rangle = \langle {\rm in}| Q_5(t_{\rm out}) |{\rm in}\rangle_{{\rm ren}} - \langle {\rm out}| Q_5(t_{\rm out}) |{\rm out}\rangle_{{\rm ren}} \, . \label{finalaux}
\eea
We can now invert this expression to finally get the result of interest:
\bea
\label{finalaux2}
 \langle {\rm in}| Q_5(t_{\rm out}) |{\rm in}\rangle_{{\rm ren}} = \langle {\rm in}| :Q_5:(t_{\rm out}) |{\rm in}\rangle+\langle {\rm out}| Q_5(t_{\rm out}) |{\rm out}\rangle_{{\rm ren}}\, .
\eea
The first contribution on the RHS depends only on the Bogoliubov coefficients and does not require renormalization. Then, it can be understood in terms of  particle  pairs created with net helicity by the external, electric  background. The second term on the RHS requires renormalization, and it represents a vacuum polarization effect.  To  the quantum anomaly, {\it both} effects contribute. 
\footnote{Mathematically,  $:Q_5:$ (and not $Q_5$) is the relevant operator that is related with the Index Theorem in geometric analysis \cite{BS16}, and, because of this, one may be tempted to identify it with the chiral anomaly. Historically, chiral anomalies were studied on compact manifolds without boundary, that arise naturally using Euclidean techniques, and in these cases the chiral fermion anomaly was found to match the predictions from the Atiyah-Singer Index Theorem. As a result, the statement that chiral anomalies are predicted by Index Theorems  became a standard lore.  However, this is not true in more general cases. In particular, for manifolds with boundary,  extra contributions arise in the index theorem, like the APS eta index $\eta_{APS}$  \cite{BS16}, and the agreement with the anomaly fails. Physically, these extra boundary terms are represented by the vacuum polarization effects $\langle {\rm out}| Q_5(t_{\rm out}) |{\rm out}\rangle_{{\rm ren}}$ pointed out in this appendix. }

As a side remark, it is interesting to note the similarity of this result with the Hawking effect for black holes. In the formation of a black hole by gravitational collapse,  one can compute the expectation value of the particle number operator using Bogoliubov transformations between past and future null infinity (with in and out states, respectively), as Hawking originally did. This calculation is well-defined and doesn't require renormalization.  In our problem, this would be analogous to the calculation of $ \langle {\rm in}| :Q_5:(t_{\rm out}) |{\rm in}\rangle$, which is related to the particle number operators. On the other hand, one can also study the Hawking effect by computing the non-diagonal, flux component of the expectation value of the stress-energy tensor across future null infinity. This calculation, based only on the in state, does require renormalization. This is because, apart from the particle pair creation, there is yet another contribution coming from vacuum polarization effects. In our case, since $Q_5$ is a quadratic operator, the evaluation of $ \langle {\rm in}| Q_5(t_{\rm out}) |{\rm in}\rangle_{{\rm ren}} $ is analogous to the calculation of the Hawking effect using the stress-energy tensor and not via the particle number operator.

The evaluation of $ \langle {\rm in}| :Q_5:(t_{\rm out}) |{\rm in}\rangle$ in (\ref{finalaux2}) is  straightforward from the expressions (\ref{op1})-(\ref{op4}) above and the canonical commutation relations. It produces:
\bea
\langle {\rm in}| :Q_5:(t_{\rm out}) |{\rm in}\rangle/\hbar & = & \sum_{\substack{n>-q\,CS[A]\\ n'<0}} |\tilde\beta^+_{n'n}|^2  -\sum_{\substack{n<-q\,CS[A]\\ n'>0}} |\beta^+_{n'n}|^2 - \sum_{\substack{n<-q\,CS[A]\\n'>0}} |\tilde\beta^-_{n'n}|^2+\sum_{\substack{n>-q\,CS[A]\\n'<0}} |\beta^-_{n'n}|^2 \, . \nonumber
\eea
Note that each sum is convergent because each summand decays in both indices as $n\to\infty$.
Using (\ref{id1})-(\ref{id2}), (\ref{id3})-(\ref{id4}) one can write: 
\bea 
\sum_{n>0}^{k}1 & = & \sum_{n>0}^{k}\left(\sum_{n''>-q\,CS[A]}^{+\infty} | \alpha^+_{n n''}|^2 + \sum_{n''<-q\,CS[A]}^{-\infty}  |\beta^+_{n n''}  |^2 \right)\, , \label{sum1}\\ 
\sum_{n<0}^{-k} 1 & = & \sum_{n<0}^{-k}\left(\sum_{n''<-q\,CS[A]}^{-\infty} | \alpha^-_{n n''}|^2 + \sum_{n''>-q\,CS[A]}^{+\infty}  |\beta^-_{n n''}  |^2  \right)\, , \label{sum2}\\
\sum_{n>-q\,CS[A]}^{k} 1 & = &\sum_{n>-q\,CS[A]}^{k}\left(  \sum_{n''>0}^{+\infty} | \alpha^+_{n'' n}|  + \sum_{n''<0}^{-\infty} |\tilde  \beta^+_{n'' n} |^2\right) \, ,\label{sum3}\\
\sum_{n<-q\,CS[A]}^{-k} 1 & = &\sum_{n<-q\,CS[A]}^{-k}  \left(\sum_{n''<0}^{-\infty} | \alpha^-_{n'' n}|  + \sum_{n''>0}^{+\infty} |\tilde  \beta^-_{n'' n} |^2\right)\, , \label{sum4}
\eea
for some positive integer $k$.  Subtracting (\ref{sum1}) from (\ref{sum3}), and then taking the limit $k\to \infty$, we obtain $ \sum_{\substack{n>-q\,CS[A]\\ n'<0}} |\tilde\beta^+_{n'n}|^2  -\sum_{\substack{n<-q\,CS[A]\\ n'>0}} |\beta^+_{n'n}|^2=[q\,CS[A]]$, where $[\cdot]$ indicates integer part. Subtracting now (\ref{sum2}) from (\ref{sum4}), and  taking again the limit $k\to \infty$, we get $\sum_{\substack{n<-q\,CS[A]\\n'>0}} |\tilde\beta^-_{n'n}|^2-\sum_{\substack{n>-q\,CS[A]\\n'<0}} |\beta^-_{n'n}|^2=-[q\,CS[A]]$. In conclusion:
\bea
 \langle {\rm in}| :Q_5:(t_{\rm out}) |{\rm in}\rangle & = & 2\hbar \, \left[q\,CS[A] \right]\, . \label{aresult}
\eea
This is the main result of this appendix. It makes manifest that the vacuum expectation value of the Dirac chiral charge $Q_5$ at late times, (\ref{finalaux2}), receives an important contribution from particle pair creation.

The evaluation of $\langle {\rm out}| Q_5(t_{\rm out}) |{\rm out}\rangle_{{\rm ren}}$ in (\ref{finalaux2}), on the other hand, is technically more involved and requires renormalization. This can be done using the adiabatic method  \cite{FNS18,BFNSV18,FNSP18}. Since the main purpose of this appendix was to show the connection with particle pair creation, we  just give the final answer without entering into the details, which is:
\bea
\langle {\rm out}| Q_5(t_{\rm out}) |{\rm out}\rangle_{{\rm ren}}  =   2\hbar \, (  q\,  CS[A]-\,  \left[q\, CS[A] \right] )\, .
\eea
The final result reads $ \langle {\rm in}| Q_5(t_{\rm out}) |{\rm in}\rangle_{{\rm ren}}=2\hbar \, q\,  CS[A]$, and $\langle {\rm in}| Q_5(t_{\rm in}) |{\rm in}\rangle_{{\rm ren}}=0$. This agrees, precisely, with the prediction of the Adler-Bell-Jackiw anomaly. In other words, there is a non-trivial evolution of the Noether charge in the quantum theory, which violates the classical symmetry. 

As a final remark, notice that, since $(q \, CS[A]- \left[q \, CS[A] \right] ) \in [0,1[$, unless $\left[q \, CS[A] \right]$ is very small, the particle-creation contribution dominates against the vacuum polarization effect.\\ 

{\bf Example: uniform electric field.}
We address now the problem described at the end of Sec. \ref{sec2}.
When the electric background field is homogeneous, $E=E(t)$,  the Weyl equations read, for any time $t$,
\bea
\begin{array}{l}
\left(i \partial_t-i \partial_x-qA_x(t)\right) u_{+}=0\, , 
\left(i \partial_t+i \partial_x+qA_x(t)\right) u_{-}=0\, .
\end{array}
\eea
In our gauge choice we have $E(t)=\partial_t A_x(t)$, with $A_x(t\to -\infty)\to 0$ and $A_x(t\to +\infty)\to A$, for some real-valued constant $A$. The Weyl equations can be solved in full closed form. The properly normalized in modes defined above are
\bea
 f_{\pm,n}^{\text {in }}(t, x) & =&\frac{1}{\sqrt{L}} e^{\mp i\left( \frac{2\pi n}{L}(t\pm x)+ q\int_{-\infty}^{t} A_x(t^{\prime}) d t^{\prime}\right)} \, ,
\eea
while the  out modes are
\bea
 f_{\pm,n}^{\text {out }}(t, x) & =&\frac{1}{\sqrt{L}} e^{\mp i\left( \frac{2\pi n}{L}(t\pm x)-q A t_{\rm out} + q\int^{t}_{t_{\rm out}} A_x(t^{\prime}) d t^{\prime}\right)} \, ,
\eea
Notice that the in modes satisfy the initial condition $f_{\pm,n}^{\text {in }}(t, x)  \sim \frac{1}{\sqrt{L}} e^{\mp i \frac{2\pi n}{L}(t\pm x)}$ at early times, while the out modes satisfy the required final condition $f_{\pm,n}^{\text {out}}(t, x)  \sim \frac{1}{\sqrt{L}} e^{\mp i \left[ t(\frac{2\pi n}{L}+q A) \pm\frac{2\pi n}{L} x \right]}$ at late times.

Using these explicit solutions, we can now calculate the relevant Bogoliubov coefficients in full closed form. It is straightforward to get
\bea
 ( f_{\pm,n'}^{\text {out }} , f_{\pm,n}^{\text {in }}) = \frac{1}{L} e^{\mp i q (At_{\rm out}+\int_{-\infty}^{t_{\rm out}} dt' A_x(t'))}  \int_0^L dx e^{\mp i\frac{2\pi}{L}(t\pm x)(n-n')}= e^{\mp i q (At_{\rm out}+\int_{-\infty}^{t_{\rm out}} dt' A_x(t'))} \delta_{n n'}\, . \label{phasefactor}
\eea
Note that (i) the result is time-independent, as expected from the Dirac inner product, (ii) $| ( f_{\pm,n'}^{\text {out }} , f_{\pm,n}^{\text {in }}) |=1$, and (iii) the emergence of $\delta_{n n'}$ is a direct consequence of the homogeneity of the background field. 

Let us assume that $q A>0$ (similar results hold when $qA<0$). Then from (\ref{positivein+})-(\ref{negativein-}) we can simplify (we neglect the irrelevant phase factor of (\ref{phasefactor})):
\bea
n>0 \, \, ({\rm positive\,\,  energy}): \quad && f^{{\rm in}}_{+, n}(t,x)  = f^{{\rm out}}_{+, n}(t,x) \, \\
n<0  \, \, ({\rm negative\,\,  energy}): \quad && f^{{\rm in}}_{+, n}(t,x)  =  \Theta(-n-qAL/2\pi) f^{{\rm out}}_{+, n}(t,x)+  \Theta(n+qAL/2\pi) {f^{{\rm out}}_{+, n}(t,x)} \, ,\\
 n<0  \, \, ({\rm positive\,\,  energy}):  \quad&& f^{{\rm in}}_{-, n}(t,x)  =  \Theta(-n-qAL/2\pi) f^{{\rm out}}_{-, n}(t,x)+  \Theta(n+qAL/2\pi) {f^{{\rm out}}_{-, n}(t,x)} \, ,\\
 n>0  \, \, ({\rm negative\,\,  energy}):  \quad&& f^{{\rm in}}_{-, n}(t,x)  =   {f^{{\rm out}}_{-, n}}(t,x) \, .
\eea
This result shows that  there is a number  $[qAL/2\pi]$ of negative-frequency, left-moving in modes that transform into positive-frequency, left-moving out modes. This means that the electric field has created $[qAL/2\pi]$ left-moving fermions (with negative electric charge) out of the in vacuum.  Similarly, there is a number   $[qAL/2\pi]$ of positive-frequency, right-moving in modes that are measured as negative-frequency, right-moving  modes by out observers. In other words, the electric field has excited $[qAL/2\pi]$ right-moving anti-fermions (with positive electric charge). All  these particles have positive helicity $\hbar$. This explains the net helicity found in the general result (\ref{aresult}).

\section{Proof of (\ref{final2})} \label{appendixIR}

We include here the technical details and computations of Section \ref{sec4}. The starting point is equations (\ref{equationphi2}) and (\ref{equationA}). Our task is to plug these expressions in (\ref{electromagnetism}) and get (\ref{final2}).

Let us denote by $A_2^{aux}\left(u, \theta, \phi\right)$ the first term on the RHS of (\ref{equationA}). Using (\ref{electromagnetism}), we divide the calculation in three terms. First:
\bea
 \int_{-L / 2}^{L / 2} d u \int d \mathbb S^2 &&\operatorname{Im}\left(A_2^{\text {aux }} \bar\Phi_2^0\right) \\
&=&\int d \mathbb S^2 \operatorname{Im} \frac{1}{L}\left\{\sum_{n,n' \neq 0} \tilde\Phi_2^0(n, \theta, \phi) \overline{\tilde\Phi_2^0}(n', \theta, \phi) \frac{e^{-i (\omega_n-\omega_{n'}) u}-e^{-i \omega_n u_0}e^{i \omega_{n'} u}}{-i \omega_n}(-1)^{n+n'}   \right\}  \nonumber \\
&+&\left.\sum_{n \neq 0} \tilde\Phi_2^0(n, \theta, \phi) \frac{e^{-i \omega_n u}-e^{-i \omega_n u_0}}{-i \omega_n} \overline{\tilde\Phi_2^0(0, \theta, \phi)}(-1)^n\right\} \nonumber\\
&=&\int d \mathbb S^2 \operatorname{Im} \frac{1}{L}\left\{\sum_{n \neq 0} L \frac{\left| \tilde\Phi_2^0(n, \theta, \phi) \right|^2}{-i \omega_n}+\frac{1}{i} L \overline{\tilde{\Phi}_2(0, \theta, \phi)} \cdot \sum_{n \neq 0} \tilde\Phi_2(n, \theta, \phi) \frac{e^{-i \omega_n u_0}}{\omega_n}(-1)^n\right\} \nonumber\\
&=&\int d \mathbb S^2 \sum_{n \neq 0} \frac{\left|\tilde\Phi_2(n, \theta, \phi)\right|^2}{\omega_n}-\operatorname{Re}\left(\overline{\tilde\Phi_2(0, \theta, \phi)} \sum_{n \neq 0} \tilde\Phi_2(n, \theta, \phi) \frac{e^{-i w_n u_0}}{\omega_n}(-1)^n\right)\, ,\nonumber 
\eea
where in the second equality we used the orthonormal properties of the basis modes (\ref{ortho}), and the identity $\int_{-L/2}^{L/2} du\, e^{i2\pi n u/L}= 0$ to get rid of some terms. Second:
\bea
 \int_{-L/2}^{L/2}du \int d \mathbb S^2 && \operatorname{Im}\left\{\left(\frac{-u_0}{\sqrt{L}} \tilde\Phi_2(0, \theta, \phi)+A_2^0\left(u_0, \theta, \phi\right)\right)\left(\sum_{n\neq 0}\overline{\tilde\Phi_2}(n,\theta,\phi)\frac{e^{i\omega_n u}(-1)^n}{\sqrt{L}}+ \frac{1}{\sqrt{L}} \overline{\tilde\Phi_2(0, \theta, \phi)}  \right)\right\}\\
&=&L  \int d \mathbb S^2 \operatorname{Im}\left\{\left(\frac{-u_0}{\sqrt{L}} \tilde\Phi_2(0, \theta, \phi)+A_2^0\left(u_0, \theta, \phi\right)\right) \frac{1}{\sqrt{L}} \overline{\tilde\Phi_2(0, \theta, \phi)}  \right\}\nonumber\\
&=&\sqrt{L} \int d \mathbb S^2 {\rm Im} \left(A_2^0\left(u_0, \theta, \phi\right) \overline{\tilde\Phi_2(0, \theta, \phi)}\right)=\int_{-L / 2}^{L/2}du d \mathbb S^2 {\rm Im} \left(A_2^0\left(u_0, \theta, \phi\right) \overline{\Phi_2\left(u, \theta, \phi\right)}\right)\, , \nonumber
\eea
where in the first equality we used the identity $\int_{-L/2}^{L/2} du\, e^{i2\pi n u/L}= 0$; in the second equality we noticed that one term was real, so its imaginary part vanishes; and in the last equality we recalled the definition of Fourier transform. Finally,
\bea
 \int_{-L / 2}^{L / 2} d u \int d\mathbb S^2 &&{\rm Im} \left\{\frac{u}{\sqrt{L}} \tilde\Phi_2(0, \theta, \phi)\left(\sum_{n \neq 0} \overline{\tilde\Phi_2\left(\omega_n ,\theta, \phi\right)} \frac{e^{+i w_n u}}{\sqrt{L}}(-1)^n+\frac{1}{\sqrt{L}} \overline{\tilde\Phi_2(0, \theta, \phi)}\right)\right\} \\
&=&\int_{-L / 2}^{L(2} d u \int d \mathbb S^2 {\rm Im} \left\{\frac{u}{\sqrt{L}} \tilde\Phi_2(0, \theta, \phi) \sum_{n \neq 0} \overline{\tilde\Phi_2\left(\omega_n \theta, \phi\right)} \frac{e^{i \omega_n u}}{\sqrt{L}}(-1)^n\right\} \nonumber\\
&=&\int_{-L / 2}^{L(2} d u \int d \mathbb S^2 {\rm Im} \left\{\frac{1}{L} \tilde\Phi_2(0, \theta, \phi) \sum_{n \neq 0} \overline{\tilde\Phi_2\left(\omega_n \theta, \phi\right)} e^{i \omega_n u}(-1)^n \frac{1}{i} \left.\frac{d}{d\epsilon}\right|_{\epsilon=0} e^{i\epsilon u}\right\}\nonumber \\
&=& \int d \mathbb S^2 {\rm Im} \left\{\frac{1}{L} \tilde\Phi_2(0, \theta, \phi) \sum_{n \neq 0} \overline{\tilde\Phi_2\left(\omega_n \theta, \phi\right)} (-1)^n \frac{1}{i} \left.\frac{d}{d\epsilon}\right|_{\epsilon=0} \int_{-L / 2}^{L(2} d u e^{i(\epsilon+\omega_n) u}\right\} \nonumber\\
&=&  \int d \mathbb S^2 {\rm Im} \left\{\frac{1}{L} \tilde\Phi_2(0, \theta, \phi) \sum_{n \neq 0} \overline{\tilde\Phi_2\left(\omega_n \theta, \phi\right)}\frac{L}{i\omega_n}  \right\} \nonumber\\
&=&  -\int d \mathbb S^2 {\rm Re} \left\{ \tilde\Phi_2(0, \theta, \phi) \sum_{n \neq 0} \frac{\overline{\tilde\Phi_2\left(\omega_n \theta, \phi\right)}}{\omega_n}  \right\}\, , \nonumber
\eea
where in the first equality we noted that one term has vanishing imaginary part; and in the fourth equality we used $ \int_{-L / 2}^{L(2} d u e^{i(\epsilon+\omega_n) u}=(-1)^n\frac{e^{i\epsilon L/2}-e^{-i\epsilon L/2}}{i(\omega_n+\epsilon)}$. 

Combining all three terms above, and for $u_0=-L/2$, which allows to further simplify some terms, we end up with
\bea
&&\int_{-L / 2}^{L / 2} d u \int d\mathbb S^2 \operatorname{Im}\left\{\left(A_2^0-\bar \eth \alpha_0\right) \bar\Phi_2\right\}\label{eleccion}\\
&=&\int d \mathbb S^2 \left\{ \sum_{n \neq 0} \frac{\left|\tilde\Phi_2(n, \theta, \phi)\right|^2}{\omega_n}-2\operatorname{Re}\left(\overline{\tilde\Phi_2(0, \theta, \phi)}   \frac{\tilde\Phi_2(n, \theta, \phi)}{\omega_n}\right) \right\} +\int_{-L / 2}^{L / 2}du d \mathbb S^2 {\rm Im} \left(A_2^0(-L / 2, \theta, \phi)-\eth \alpha_0\right) \overline{ \Phi_2(u, \theta, \phi)}\, . \nonumber
\eea
Note that $\sum_{n\neq 0}\frac{\left|\tilde\Phi_2^0(0, \theta, \phi)\right|^2}{\omega_n}\sim \sum_{n\neq 0}\frac{1}{n}=0$. Therefore, we can rewrite the first term above  as $\int d \mathbb S^2 \sum_{n \neq 0} \frac{\left|\tilde\Phi_2^0(n, \theta, \phi)-\tilde\Phi_2^0(0, \theta, \phi)\right|^2}{\omega_n}$. The second term above can also be greatly simplified and has a nice physical interpretation. To see this, note first that, in the gauge $A_1^0=A_a n^a =0$, the 1-form $A_a^0$ lives on the tangent space of  $\mathbb S^2$, so it can be expressed as the sum of a gradient and a curl: $A_a^0=D_a \alpha+\epsilon_a^{\hspace{0.15cm}b}D_b\beta$, where $D_a$ is the connection on $\mathbb S^2$. Thus, $A_2^0=\bar\eth(\alpha+i\beta)$, with $\alpha,\beta\in C^{\infty}(\mathbb R\times \mathbb S^2,\mathbb R)$. There still exists a residual gauge freedom represented by $A_a^0 \to A_a^0+D_a \Lambda$, with $\dot \Lambda=0$. Under this transformation $\beta$  remains invariant, and although $\alpha$ doesn't remain invariant, the combination $\alpha-\alpha_0$ does. With this new terminology we can reexpress the second term on the RHS above as
\bea
\int_{-L / 2}^{L / 2}du d \mathbb S^2 {\rm Im} \left(A_2^0(-L / 2, \theta, \phi)-\eth \alpha_0\right) \overline{ \Phi_2^0(u, \theta, \phi)} &=& \int_{-L / 2}^{L / 2}du d \mathbb S^2 {\rm Im} \left(i\bar \eth \beta+\bar \eth (\alpha-\alpha_0)\right) \overline{ \Phi_2^0(u, \theta, \phi)}\nonumber\\
&=& -\int_{-L / 2}^{L / 2}du d \mathbb S^2 \beta{\rm Re}  \overline{ \eth \Phi_2^0(u, \theta, \phi)}+(\alpha-\alpha_0)\beta{\rm Im}  \overline{ \eth \Phi_2^0(u, \theta, \phi)}\nonumber\\
& =& -{\rm Re} q_{\beta} + {\rm Im}q_{\alpha-\alpha_0}\, ,
\eea
where in the last step we made use of Maxwell equations (\ref{maxwell1}) and the definition of infrared charges (\ref{softcharges}).

As we can see this contribution in the chiral anomaly is related to  memory of the electromagnetic background. Despite $\Phi_2^0(\pm\frac{L}{2},\theta,\phi)=0$, if $\Phi_2^0(u,\theta,\phi)\neq 0$, then $\beta(u,\theta,\phi)$ and $\alpha(u,\theta,\phi)-\alpha_0(u,\theta,\phi)$ evolve in time $u$, leading to non-zero soft charges. Using (\ref{imphi1})-(\ref{rephi1}), these functions can be solved from ${\rm Re} \,\Phi_1^0(u,\theta,\phi)=\Delta \alpha(u,\theta,\phi)+G(\theta,\phi)$ and ${\rm Im}\, \Phi_1^0(u,\theta,\phi)=\Delta \beta(u,\theta,\phi)$, where $\Delta$ denotes the 2-dimensional Laplacian. The parameter $\beta(-L/2,\theta,\phi)$ tells us about the magnetic field at spatial infinity $i^0$ when we take $L\to \infty$, while the combination $\alpha(-L/2,\theta,\phi)-\alpha_0(-L/2,\theta,\phi)$ tells us about the electric field. The residual gauge freedom mentioned above can be fully fixed by choosing $G$ such that $G(\theta,\phi)=-\frac{Q}{2}$, where $Q=\frac{1}{2\pi}\int_{\mathbb S^2}d\mathbb S^2 {\rm Re} \,\Phi_1^0$ is the (constant) electric charge of the sources. Because $\alpha_0$ was chosen such that $A_a^0=D_a \alpha_0$ produces $\Phi_2^0=\Phi_1^0=0$, in this fully fixed gauge choice $\alpha_0$ satisfies $\Delta \alpha_0=0$, which for smooth functions on the sphere is equivalent to $\alpha_0=0$. In summary, for a fully gauge-fixed theory the final result reads:
\bea
\int_{-\infty}^{\infty} d u \int d\mathbb S^2 \operatorname{Im}\left\{\left(A_2^0-\bar\eth \alpha_0\right) \bar\Phi_2^0\right\} = \frac{1}{2\pi}\int d \mathbb S^2  \int_{-\infty}^{\infty} d\omega \frac{\left|\tilde\Phi_2^0(\omega, \theta, \phi)-\tilde\Phi_2^0(0, \theta, \phi)\right|^2}{\omega} + {\rm Im}q_{\alpha} -{\rm Re} q_{\beta} \, , \label{final}
\eea
where the infrared charges are evaluated for $\alpha(-\infty,\theta,\phi)$, $\beta(-\infty,\theta,\phi)$.

Note that the integral is well-defined in the infrared limit $\omega \to 0$. On the other hand, when the soft charges are all zero, we also have $\tilde\Phi_2^0(0, \theta, \phi)=0$, and this expression reduces to the result obtained in \cite{dR21}.  As shown in \cite{dR21}, the first contribution on the RHS above represents the difference between right-handed and left-handed circularly polarized radiation reaching future null infinity. This is a purely radiative contribution.

The contribution from the IR charges in (\ref{final}) can be rewritten in a different form:
\bea
{\rm Im}q_{\alpha} -{\rm Re} q_{\beta} & = & \int d\mathbb S^2 \alpha(-\infty,\theta,\phi)({\rm Im} \Phi_1^0(\infty,\theta,\phi)-{\rm Im} \Phi_1^0(-\infty,\theta,\phi)) -\int d\mathbb S^2 \beta(-\infty,\theta,\phi)({\rm Re} \Phi_1^0(\infty,\theta,\phi)-{\rm Re} \Phi_1^0(-\infty,\theta,\phi))\nonumber \\
 & = & \int d\mathbb S^2 \alpha(-\infty,\theta,\phi)\Delta (\beta(\infty,\theta,\phi)- \beta(-\infty,\theta,\phi)) -\int d\mathbb S^2 \beta(-\infty,\theta,\phi) \Delta ( \alpha(\infty,\theta,\phi)-\alpha(-\infty,\theta,\phi))\nonumber\\
 & = & \int d\mathbb S^2 \left[\alpha(-\infty,\theta,\phi)\Delta \beta(\infty,\theta,\phi) -  \beta(-\infty,\theta,\phi)\Delta \alpha(\infty,\theta,\phi) \right]\, . \label{remark}
\eea

As a final remark, note that equation (\ref{final}) can also be obtained if we had chosen $u_0=+L/2$ to write (\ref{eleccion}) instead of $u_0=-L/2$. The only difference is that $\alpha$ and $\beta$ in that equation would be evaluated at $u=+L/2$ and not at $u=-L/2$. Although apparently different, the two expressions are actually equivalent. The equivalence is manifest from  (\ref{remark}), which implies ${\rm Im}q_{\alpha(\infty)} -{\rm Re} q_{\beta(\infty)}={\rm Im}q_{\alpha(-\infty)} -{\rm Re} q_{\beta(-\infty)}$ after integration by parts.  Equation (\ref{final}) with this last choice is the result of equation (\ref{final2}) of the main text.

\end{document}